# Quantum Nanophotonics in Two-Dimensional Materials

Antoine Reserbat-Plantey[1], Itai Epstein[1,2], Iacopo Torre[1], Antonio T. Costa[3], P. A. D. Gonçalves[4], N. Asger Mortensen[4,5,6], Marco Polini[7,8,9], Justin C. W. Song[10,11], Nuno M. R. Peres[3,12], Frank H. L. Koppens[1,13]

[1]ICFO-Institut de Ciencies Fotoniques, The Barcelona Institute of Science and Technology, 08860 Castelldefels (Barcelona), Spain
[2] Department of Physical Electronics, School of Electrical Engineering, Tel Aviv University, Tel Aviv 6997801, Israel
[3]International Iberian Nanotechnology Laboratory, 4715-330 Braga, Portugal
[4]Center for Nano Optics, University of Southern Denmark, DK-5230 Odense M, Denmark
[5]Danish Institute for Advanced Study, University of Southern Denmark, DK-5230 Odense M, Denmark
[6]Center for Nanostructured Graphene, Technical University of Denmark, DK-2800 Kgs. Lyngby, Denmark
[7]Dipartimento di Fisica dell'Università di Pisa, Largo Bruno Pontecorvo 3, I-56127 Pisa, Italy
[8]School of Physics & Astronomy, University of Manchester, Oxford Road, Manchester M13 9PL, United Kingdom
[9]Istituto Italiano di Tecnologia, Graphene Labs, Via Morego 30, I-16163 Genova, Italy
[10]Division of Physics and Applied Physics, Nanyang Technological University, 637371, Singapore
[11]Institute of High Performance Computing, Agency for Science, Technology, and Research, 138632, Singapore
[12] Centro de Física das Universidades do Minho e Porto and Departamento de Fısica and QuantaLab, Universidade do Minho, Campus de Gualtar, 4710-057 Braga, Portugal
[13]ICREA – Institució Catalana de Recerca i Estudis Avançats, 08000 Barcelona, Spain

Corresponding authors: antoine.reserbat-plantey@icfo.eu - Frank.koppens@icfo.eu

**Abstract**

The field of 2D materials-based nanophotonics has been growing at a rapid pace, triggered by the ability to design nanophotonic systems with in-situ control[1], unprecedented degrees of freedom, and to build material heterostructures from bottom up with atomic precision[2]. A wide palette of polaritonic classes[3–6] have been identified, comprising ultra-confined optical fields, even approaching characteristic length-scales of a single atom[7]. These advances have been a real boost for the emerging field of quantum nanophotonics, where the quantum mechanical nature of the electrons and/or polaritons and their interactions become relevant. Examples include, quantum nonlocal effects[8–11], ultrastrong light–matter interactions[11–16], Cherenkov radiation[13,17,18], access to forbidden transitions[11], hydrodynamic effects[19–21], single-plasmon nonlinearities[22,23], polaritonic quantization[24], topological effects etc.[3,4]. In addition to these intrinsic quantum nanophotonic phenomena, the 2D material system can also be used as a sensitive probe for the quantum properties of the material that carries the nanophotonics modes, or quantum materials in its vicinity. Here, polaritons act as a probe for otherwise invisible excitations, e.g. in superconductors[25], or as a new tool to monitor the existence of Berry curvature in topological materials and superlattice effects in twisted 2D materials.

In this article, we present an overview of the emergent field of 2D-material quantum nanophotonics, and provide a future perspective on the prospects of both fundamental emergent phenomena and emergent quantum technologies, such as quantum sensing, single-photon sources and quantum emitters manipulation. We address four main implications (cf. **Figure *1***): i) quantum sensing, featuring polaritons to probe superconductivity and explore new electronic transport hydrodynamic behaviours, ii) quantum technologies harnessing single-photon generation, manipulation and detection using 2D materials, iii) polariton engineering with quantum materials enabled by twist angle and stacking order control in van der Waals heterostructures and iv) extreme light–matter interactions enabled by the strong confinement of light at atomic level by 2D materials, which provide new tools to manipulate light fields at the nano-scale (e.g., quantum chemistry[26], nonlocal effects, high Purcell enhancement).

**Keywords**: 2D materials, quantum photonics, light-matter interactions, polaritons, single photon

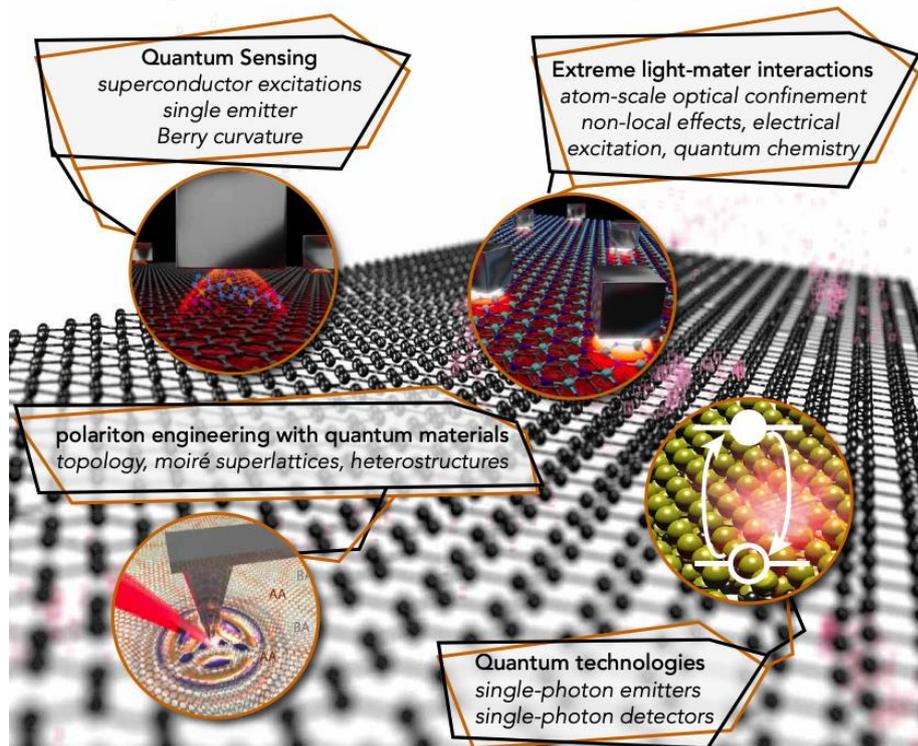

**Figure 1. Perspectives in quantum nanophotonics with 2D materials.** The co-evolution of subwavelength optical measurements and photonic devices designed atoms by atoms led to the exploration of new regimes where strong light matter interactions enable original strategies of quantum sensing, polariton engineering and quantum technologies.

| Glossary | |
|---|---|
| **AGP** | acoustic graphene plasmons |
| **FQH** | fractional quantum Hall |
| **GDM** | graphene–dielectric–metal |
| GP | graphene plasmons |
| **GPMR** | graphene-plasmon-magnetic-resonators |
| **hBN** | hexagonal Boron-Nitride |
| **NPoM** | nanoparticle-on-metal |
| **MIR** | Mid infra-red |
| **s-SNOM** | scattering-type scanning near-field optical microscopy |
| **SC** | Superconductor |
| **SPP** | surface plasmon polaritons |
| **TBMs** | topological boundary modes |
| TMD | transition-metal-dichalcogenides |

**Pushing nanophotonics to the ultimate physical limits**

2D materials provide a path for driving nanophotonics beyond the nanometer scale and into the atomic scale. This naturally stems from their atomically thin nature, but also from the unique optical excitations supported by these materials[3,4], such as plasmons in graphene[12,27–29], excitons in TMDs[30], etc. Although similar quasi-particle excitations exist in conventional bulky materials, the low-dimensionality and unique composition of 2D materials yields unconventional optical properties and enhanced quantum effects.

There are several limiting factors for controlling light at the nanoscale, with the primary one being the ability to compress the optical field to small volumes. The latter has an important effect not only on the accompanied field enhancement but also on the Purcell factor, which allows the manipulation of single quantum emitters. This challenging task becomes even more difficult below the nanometer scale, due to the increased momentum mismatch between the far-field photon and the excited optical mode, together with enhanced quantum effects governing this regime. However, overcoming these challenges provides a path to ultrastrong and enhanced light–matter interactions. In that aspect, 2D materials and their properties enabled pushing light–matter interaction into the atomic limit, based on their unique optical response. For example, the quantum response of excitons in monolayer semiconductor transition-metal-dichalcogenides (TMDs) enabled the realization of highly reflective mirrors[31,32] and perfect absorbers[33,34], which are one monolayer thick. Furthermore, the relative ease of controlling the excitonic properties of these materials enabled the large manipulation of their radiative lifetime via optical cavities[35–38], together with quantum nonlinear effects at the single photon level, which provide a possible path for new platforms based on strongly interacting photons[32,33,39–42]. Similarly, finite-sized graphene structures supporting graphene plasmons (GPs) give rise to quantum effects and single-plasmon response[23,43,44].

One exemplary system for light confinement is the nanoparticle-on-metal (NPoM) system, which is composed out of a nanometric size metal particle (usually a cube or a sphere), separated from a metallic surface by a very thin dielectric spacer[45,46] (**Figure 2**a). The NPoM allows to confine a gap surface-plasmon mode between the nanoparticle and the metal surface to a small mode volume with large field enhancement (~$10^4$) and Purcell factor (~$10^5$-$10^7$)[47,48]. It has thus enabled room-temperature strong-coupling[49], spontaneous emission sources[50] and nanoantenna response[45,49]. However, the confinement provided by the NPoM system is also accompanied by increased losses due to its plasmonic nature[51,52], and an important role in the reduction in the possible field enhancement is introduced by quantum nonlocal effects[53–55] (see **Nonlocality** section).

Taking the concept of the NPoM to the next level has been enabled by 2D materials. From a technical aspect, these allow to build NPoM systems with atomic layer-by-layer precision, thinning down the dielectric spacer to a single monolayer[56,57] (**Figure 2**b). More fundamentally, harnessing the properties of GPs, which in comparison to conventional (metal-based) SPPs exhibit extreme confinement and low loss in the MIR range[58], enabled to strongly compress even the long wavelength MIR spectrum. By replacing the metal surface with a graphene sheet, the supported GPs interacting with the metallic nanocube are able to form localized graphene-plasmon-magnetic-resonators (GPMRs), and achieve a confinement factor ~$5.10^{-10}$ time smaller than that of the free space photon volume[59] (**Figure 2**c). Moreover, even if graphene is placed at a distance of one atomic monolayer from a metallic grating, GPs can be vertically confined to that one atomic layer spacing, without experiencing prohibitive losses[57], which constitutes a tremendous advantage compared to the SPP based system. It stems from the unique nonlocal response of graphene, which do not increase the losses or limit the field enhancement[60], together with the low metallic losses in the MIR. These effects will be further discussed in **Nonlocality** section.

The ultra-compressed mode volumes provided by the GPMR system has a tremendous impact on light-emitter interaction via the Purcell effect. The latter affects the transition rates of the emitter and can be manipulated via the environment of the emitter, changing the local density of photonic states. Furthermore, GPs have been predicted to enable forbidden optical transitions[13], manipulate the frequencies of semiconductor emitters[61], the quantum Cherenkov effect[18] and the tailoring of various types of light–matter interactions[14]. In addition to the Purcell effect, these extraordinary properties makes the GPMR system an excellent candidate for pushing light-matter interaction into the ultrastrong regime[14]. Due to the small GP wavelength and mode volume, the light field intensity is significantly increased. Together with the fact that these resonances reside in the MIR and THz spectra, it makes GPs promising candidates for molecular sensing[62,63], as many molecular transitions resides in this spectral range. Moreover, this large field intensity may potentially yield very strong interaction with a nearby molecule, paving a path to ultrastrong vibrational-strong-coupling[64,65], and even the polaritonic manipulation of chemical processes[26,66].

The ultra-compressed mode volumes provided by 2D photonic materials are facilitating strong mutual coupling of quantum emitters and the large variety of photonic quasi-particles that can be hosted by 2D materials[14]. This paves the way for new paradigms in light-matter interactions beyond the Wigner-Weisskopf regime, such as strong-coupling phenomena[67], breaking of common selection rules by strong field gradients[13], and new non-perturbative regimes of QED[68]. Naturally, this development calls for new theoretical foundations, that treat light, matter, and their mutual couplings at the same footing when accounting for quantization. Examples of such developments include variational theory of general non-relativistic QED systems of coupled light and matter[69] and harnessing light emission through the exploitation of vacuum forces in novel nanophotonic systems[70].

Another advantageous attribute of GPs is that they are electrically tunable by changing the density of charge carriers in graphene. This enables to electrically control polaritonic phenomena, and indeed GP-based electro-optical detectors[71] and modulators[72] have been demonstrated. The next step in to the quantum regime is the electrical excitation of GPs by an atomic-scale quantum tunneling device, which had also been proposed[73,74]. The tunneling in such a device occurs between two graphene layers and a few Angstroms thick hBN barrier, which is controlled with atomic precision[73,74] (**Figure 2**d). Its realization would enable a complete and compact electro-optical quantum system in the MIR/THz range.

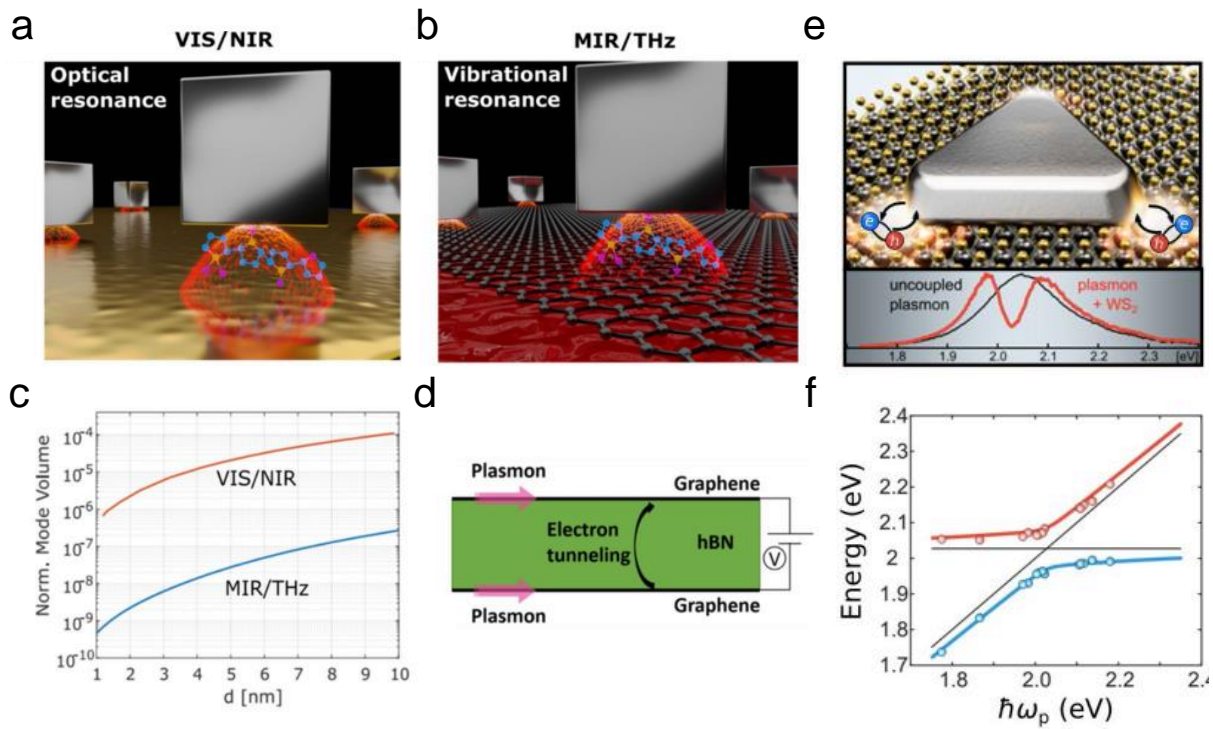

**Figure 2 : Approaches for nanophotonic cavities with confined optical modes. a**: The nanocube-on-metal (NCoM) system, which support a gap SPP in the VIS\NIR confined between a metal nanocube and a metal surface, interacting with an optical emitter. **b**: The graphene-plasmon-magnetic-resonator (GPMR) system, which support a GP in the MIR\THz confined between a metal nanocube and a graphene sheet, interacting with a vibrational molecular resonance. **c**: The normalized mode volume $V_{mode}/V_{free-space}$ for the two systems, showing a four orders of magnitudes smaller values for the GPMR system. **d**: A graphene tunneling device, where a few layers of hBN separate to graphene layers, which are electrically connected and can excite GPs via the tunneling electrons. **e**, Illustration of the coupling between excitons in a TMD and the plasmonic field of a metal nanoparticle (top), which leads to double-peaked spectra (bottom). **f**, Anticrossing obtained through dark-field microscopy of several Ag nanoprisms on monolayer $WS_2$. **e** and **f** are adapted from Ref.[75].

Strong light-matter interactions in 2D materials can also be explored with exciton-polaritons[3,4,15,76], which become relevant when the coupling rate between the optical mode and the the exciton exceeds any dissipation rate of the system[15,77,78] (cf. **Figure 2**e-f). Strong exciton–photon coupling has been experimentally demonstrated with different TMDs in both dielectric-based[76,79–82] and plasmonic-based[75,83–88] optical cavities (cf. **Figure 2**e). In spite of these early encouraging developments, the study of strong-coupling phenomena with TMDs still faces some challenges. For instance, the majority of the strong-coupling claims squarely rely on the observation of mode splitting (cf. **Figure 2**f) in the scattering spectra which is insufficient to determine whether or not the system is in the strong-coupling regime, as both Fano-type interference/induced-transparency or enhanced absorption can yield similar signatures[15,16,77,89–91]. To distinguish strong-coupling from these, it has been suggested that scattering data should be supplemented by the observation of a doublet in absorption and/or photoluminescence as well[76,86,89–92]. Ideally, the "smoking-gun" would be the observation of Rabi-like oscillations in the time-domain (though experimental challenging—but possible[93] —due to the fast (~ 10 fs) oscillation time scales), signaling a coherent, reversible exchange of energy between the exciton and photon modes. Furthermore, most of the experimental observations of strong-coupling in TMD-based systems could so far be well-described semiclassically using, e.g., the celebrated coupled-oscillator model[94]. While this is not a problem per se for many applications (e.g., for controlling emission/absorption dynamics or tailoring emission patterns), for genuine quantum light applications it is paramount to distinguish if the nature of the strong coupling is classical or quantum. This can be elucidated, e.g., by analyzing photon statistics and correlations[95,96] or few-photon nonlinearities[97–99]. Strong-coupling nanophotonics with 2D TMDs is a fertile research ground with many promising opportunities on the horizon, such as polaritonic lasing[100,101] and condensation[102,103], taming chemical reactions via polaritonic chemistry[104,105], all-optical switching and logic[98,106], and valley-polaritonics[80,107,108].

**Single quantum emitters coupled to 2D materials.**

Placing single emitters in the near-field of a surface and controlling their emission by either tuning the emitter-surface separation or the optical conductivity of the 2D material (e.g., via electrostatic gating), calls for fundamental physics questions about dipolar interactions[109] such as Casimir forces[110] or energy transfer [109,111–114]. One can see the radiative decay rate as an extremely short-range probe (interactions scaling as $d^{-4}$, where d is the separation (cf. **Figure 3**a) to quantitatively measure distances[111,112], superconductors phase transitions[115], exciton-polariton formation[116] or even dimensionality of a quantum emitter[113]. The ability to dynamically tune either the emission rate (via radio-frequency optomechanical device[117]) or directly the energy (vacuum quantum fluctuations[110]., Stark coupling[118], cf. **Figure 3**b) of a single photon source is very promising for future quantum technologies (quantum key distribution, high bandwidth quantum sensors).

Light–matter interactions between single emitters and graphene can be tuned even further by modulating the local density of states within the 2D membrane *in-situ* so that single emitters couple to graphene plasmons, enabling extremely high Purcell enhancement factors[119,120]. Demonstration of modulation faster than the emitter decay time[120] opens perspectives in exploring intriguing effects such as collective effects[121], non-linear light–matter interactions at the quantum level[22], and temporal quantum control of a single emitter[122]. Combining NV center noise magnetometry techniques with 2D electrical transport devices also provide a new tool for exploring fundamental physical phenomena locally and dynamically (current flow imaging[123], electron-phonon Cerenkov instability[17]), as illustrated in **Figure 3**d.

2D materials can also directly host single-photon sources[124–130], now integrated in devices (quantum LED[131], quantum Stark-confined modulators[132], SiN photonic chip[133]). Their emission generally ranges from visible to near-IR at low temperatures and even at room temperature for color centers in hBN. These emitters are generally well defined in energy (1-10 GHz linewidth[134,135]), emit in a broad spectrum from visible (hBN) to near-infrared (TMDs), and they are ultra-sensitive to their local environment, being located *at* the surface. Strain engineering of the membrane (nano-pillars[136], nano-constrictions[137]) or He-FIB (5 nm resolution) induced defects[138,139] are currently the main strategies to deterministically generate single-photon sources (cf. **Figure 3**e) and opens avenues in the spatial control of single emitters at subwavelength distances to engineer collective behavior and strong dipole-dipole interactions (cf. **Figure 3**f). Advanced STM electro-luminescence at low temperature has revealed atomic scale mapping of localized excitons and rises fundamental questions about the defect type and single-photon source formation[140,141] (cf. **Figure 3**g). Another promising route[142] to generate quantum emitters in 2D materials is based on the moiré superpotential to trap interlayer excitons[143,144] in $MoSe_2/WSe_2$ heterostructures[145–148]. Initial results reveal very narrow lines (<meV), showing particular magnetic dependence[146,149], have been recently competed by photon statistics[150] (anti-bunching) thus proving the existence of single photon generation. While arrays of single quantum emitters in 2D are promising tools[151] for creating Hubbard systems or exciton trapping, reaching a high level of control of single photon sources in 2D materials will open avenues for quantum communications (on demand photon sources, quantum key distribution) and quantum sensing. Such sensing capabilities are linked to the recent use of graphene plasmons to probe nonlocal effects[152].

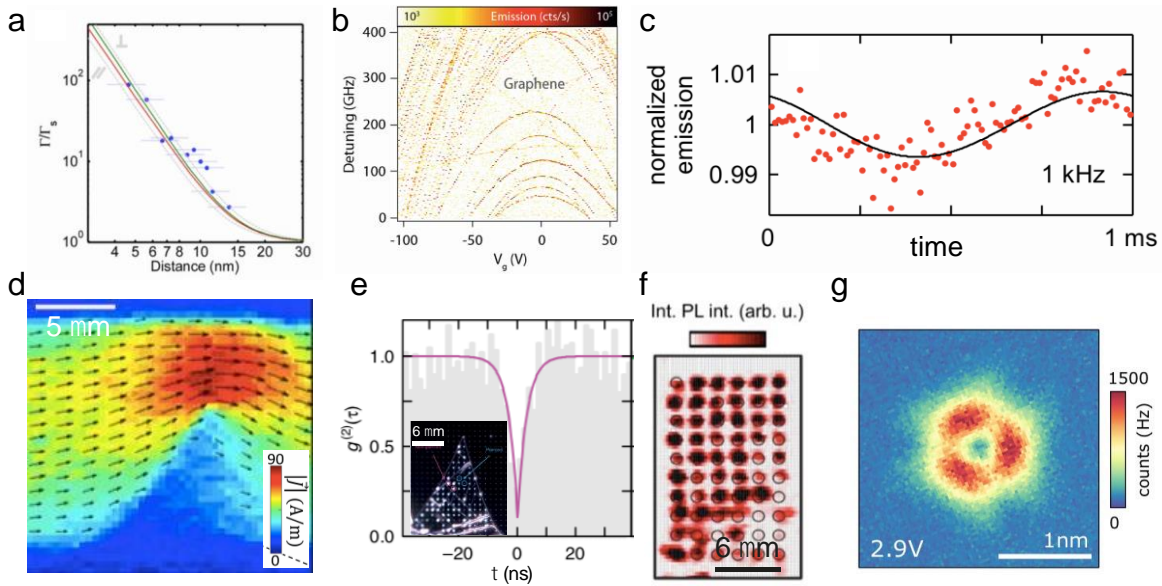

**Figure 3: Quantum emitters coupled to 2D materials. a**: Decay rate of a single emitters located at a distance d from graphene. The typical increase of the decay rate at short distance is due to non-resonant energy transfer (Ref[111]). **b**: Broadband Stark tuning of ultra-narrow (50MHz) quantum emitter using a graphene electrode. The achieved tuning is 4 orders of magnitude larger than the emitter linewidth, which is a record due to the 2D vertical geometry allowing extremely high electric field (MV/m) (Ref[118]). **c:** Dynamic modulation within the plasmon regime using graphene electrode in the near field of erbium ions (Ref[120]). **d:** current flow near a defect in graphene measured by NV-center magnetometry (Ref[123]). **e**: anti-bunching signature of a single-photon sources in WSe2 generated by strain fields using nano-pillar array (inset) (Ref[136]). **f**: Emission from defects in MoS2 caused by He ion exposure on a square pattern (Ref[139]). **g:** Photon-emission spatial imaging of a single sulfur vacancy in WSe2 measured by electrically stimulated photon emission using STM tip (Ref[153]).

**Nonlocality**

The recent realization of acoustic graphene plasmons (AGPs) in graphene–dielectric–metal (GDM) heterostructures[8,10,59,154,155], exhibiting a nearly-linear dispersion and unparalleled field confinement, revealed a new paradigm in graphene plasmonics: the ability to probe and tune quantum nonlocal effects and to simultaneously map the both the frequency- and momentum-dependence (*i.e.*, nonlocality or spatial dispersion) of graphene's conductivity, $\sigma(q,\omega)$. In general, nonlocal effects impact the electromagnetic response of materials when these are probed at wavevectors $q$ comparable with the Fermi wavevector $k_F$ of the underlying electron system, or, else, for $q \to \omega/v_F$ (signaling the breakdown of the $q \ll \omega/v_F$ condition)[152,156] corresponding to AGP's velocities, $v_{AGP} \sim \omega/q$, approaching the electron's Fermi velocity, i.e., $v_{AGP} \to v_F$. Additionally, as AGPs in GDM structures attain large wavevectors (even larger than conventional GPs) and that graphene's $k_F$ can be tuned electrostatically by controlling its carrier density $n$ through $k_F = \sqrt{\pi|n|}$, these features can be exploited to produce AGPs with wavevectors corresponding to a significant fraction of $k_F$ or with plasmon velocities approaching $v_F$.

In extended graphene, the main effect of nonlocal response is to shift the plasmon dispersion towards smaller $q$ (cf. **Figure 4**a-b). For AGPs in GDM structures, this shift can be substantial at small graphene–metal separations and is also responsible for the slowdown of the AGPs' velocity[8,10]. Moreover, as the graphene–metal separation $t$ is reduced, the nonlocal description predicts plasmon velocities asymptotically approaching electron velocity $v_F$, but without ever surpassing (falling below) it[8,10,157]. This is in stark contrast to the local-response prediction which allows the AGP's dispersion to fall inside the prohibited graphene's intraband electron-hole continuum. Remarkably, Lundeberg *et al.*[8] were not only able to detect a markedly nonlocal response, but also to uncover intriguing many-body effects, specifically, Fermi velocity renormalization and compressibility correction. This exciting development suggests that near field optical spectroscopy with AGPs can be used to investigate electron-electron interactions in (twisted) graphene[158,159].

Strikingly, the significance of AGPs goes beyond graphene itself as much as they can also be employed as extremely sensitive probes of the nonlocal and quantum response of metals[9]. While the classical electromagnetic response of metals in the far-infrared mimics that of a perfect conductor, at nanometric graphene–metal separations, $t$, that is no longer an accurate approximation as $t$ approaches intrinsically quantum mechanical length scales associated with the metal's electron gas[9,10,43]. The impact of the metal's non-classical response causes the AGP's dispersion to shift; for gold (a typical metal used in GDM structures) it is towards smaller $q$ (or larger $\omega$; blueshift) [9,10] whereas simple metals (*e.g.,* Al or Na) should induce an opposite shift. Curiously, such non-classical shift is *not* accompanied by a significant nonlocal Landau damping from the metal. Both of these features—*i.e.*, nonclassical spectral shifting and negligible metallic nonlocal broadening—can be well explained using a mesoscopic framework for nanoscale electrodynamics where the quantum surface-response of the metal is encoded via the so-called Feibelman $d$-parameters[10,160,161]: $d_\perp(\omega)$ and $d_\parallel(\omega)$. Within a simple jellium treatment[160,161], $d_\parallel = 0$. In this context, the $d_\perp$-parameter is particularly prominent as it corresponds to the first moment of the induced charge density, and thus $Re(d_\perp)$ indicates the effective position of the metal's surface with respect to its jellium edge (which marks the "classical" surface), while $Im(d_\perp)$ embodies surface-enhanced Landau damping. Therefore, the metallic quantum surface-response effectively renormalizes the graphene–metal distance[54] from $t$ to:

$$\tilde{t} = t - Re(d_\perp).  \quad \text{(Eq. 1)}$$

The magnitude and direction of the quantum shift thus depends, respectively, on the absolute value and sign of $Re(d_\perp)$ (cf. Figure 4c). For the THz and mid-IR frequency range, *i.e.*, much below the metal's plasma frequency ($\omega \ll \omega_P$) but corresponding to the spectral range of interest for AGPs, it can be shown[161,162] that:

$$d_\perp \approx Re(d_\perp) = \zeta \quad \text{(Eq. 2)}$$

where $\zeta \equiv Re(d_\perp)(\omega \to 0)$ is a constant and $Im(d_\perp)$ asymptotically vanishes as $\omega \to 0$. Hence, the finiteness of $Re(d_\perp)$ and the nearly vanishing $Im(d_\perp)$ elucidate the physical mechanisms underpinning the metal's quantum surface-response that are responsible for the occurrence of AGP's dispersion shifts without the significant deterioration of the associated quality-factor (cf. **Figure 4**c-d), and, crucially, explains recent experimental observations[155].

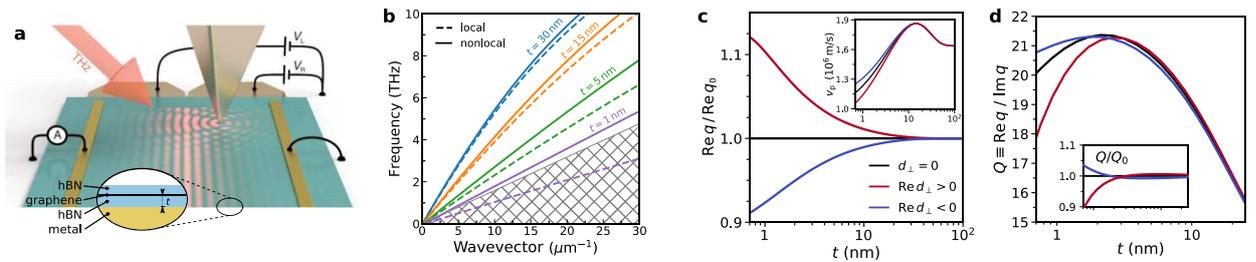

**Figure 4 : Nonlocal and quantum plasmonics in graphene–metal hybrids. a**, Typical experimental setup for launching and nanoimaging of AGPs using a s-SNOM; the sample consists of hBN–graphene–hBN–metal. Adapted from Ref.[152]. **b**, Dispersion of AGPs for varying graphene–metal distances, t, obtained from local- (Drude; dashed lines) and nonlocal-response (RPA; solid lines) theories. Setup parameters: $\epsilon^{hBN}_\perp = 6.7$, $\epsilon^{hBN}_\parallel = 3.56$, the top hBN layer is 10 nm-thick, and graphene's carrier

density and relaxation-time have been set to $n = 10^{12} cm^{-2}$ and $\tau = 500 fs$, respectively. The metal is treated classicaly (within the local-response). **c**, Impact of metallic quantum surface-response on the spectral properties of AGPs for varying t in graphene–dielectric–metal structures, contrasting the metal's response based on classical ($d_\perp = 0$) and quantum ($d_\perp = \zeta + i\xi\omega/\omega_P$, with $\zeta = 4$Å and $\xi = 1$Å) treatments. The inset shows the corresponding AGP's group velocity, $v_P = \partial\omega/\partial q$. **d**, Associated AGPs' quality factor, Q. The inset shows the ratio between the quality factors based on calculations without ($Q_0$ ; $d_\perp = 0$) and with ($Q$ ; $d_\perp \neq 0$) the metal's quantum surface-response. Setup parameters (panels **c–d**): ($r_s = 3, \hbar\gamma_m = 0.1eV, \epsilon_d = 4, E_F = 0.3eV$ and $\hbar\gamma = 8meV$; we assume an excitation at $\lambda_0 = 11.28\mu m$ or $f_0 \approx 26.6 THz$). Adapted from Ref.[163].

Another intriguing avenue is the direct probing of compressibility through plasmonic excitations in correlated systems such as the fractional quantum Hall (FQH) system or magic angle graphene[164–166]. Spatial mapping of the nonlocal conductivity in the FQH regime *via* scattering-type scanning near-field optical microscopy (s-SNOM) may shed light on the origin of incompressibility. Indeed, the nonlocal conductivity carries precious information on the local shape of correlations (pair distribution function) and on an internal geometrical degree of freedom (internal metric) responsible for the intra-Landau level dynamics of the system[167]. S-SNOM can also directly probe collective excitations in the FQH regime. It is by now well understood that, at finite wave numbers, the system supports a sharp magnetoroton mode exhibiting a "roton" minimum[168], which has been observed experimentally via inelastic light scattering[169]. On the other hand, the understanding of collective modes in the long-wavelength regime is far from being complete. The authors of Ref.[170] have recently demonstrated that the long-wavelength collective excitations of FQH liquids can be viewed as "chiral gravitons" carrying total angular momentum 2, mismatching that of the photon which carries angular momentum 1. While these gravitons can be excited and probed by using surface acoustic waves[171] (whose effects in a crystal mimic those of gravitational waves), it would be highly desirable to engineer metallic tips capable of probing them with s-SNOM. There is thus a clear perspective to exploit near-field optical spectroscopy for the study of non-local conductivity and probe electron-electron interactions and unexplored collective excitations, which can be crucial to explore electronic phases of matter.

**Collective modes in electronic hydrodynamic systems**
The long-standing and intuitive idea that conduction electrons can move like a viscous fluid in good conductors[172,173] has recently found experimental verification on graphene[19,20,174,175], PdCoO$_2$[176], WP$_2$[177], and WTe$_2$[178]. In high quality graphene layers, for example, electrons are able to travel long distances without colliding with external entities (phonons or impurities) while, at "high" electron temperatures (higher than liquid nitrogen temperatures), the electron-electron (*e-e*) scattering time is sufficiently short[19]. This makes *e-e* scattering (due to Coulomb interaction[179] or phonon-mediated interaction[180]) the dominant momentum-transfer mechanism in this regime. While, the impact of this new transport regime[20,181] has been thoroughly studied using steady-state transport and various scanning techniques, much less is known on the impact of hydrodynamic transport on optical properties (that is beyond the framework of hydrodynamic model of the nonlocal optical response of metals[182]).

Electron hydrodynamics can suggest very elegant analogies between electronic and fluid-mechanical phenomena[173,183,184]. Interestingly, recent theoretical works[185–187] showed that in the hydrodynamic regime of a Fermi liquid plasmon modes are possible with a phase velocity lower than the Fermi velocity $v_F$ (but greater than $v_F/\sqrt{2}$), as illustrated in Figure 7c. This allows to push the impact of nonlocal effects to a new limit. An even richer landscape is expected when we consider non-Fermi liquid systems[188,189]. For example, for graphene close to charge neutrality, fast *e-e* scattering, extremely low viscosity, and neutral energy modes have been predicted[185–187,189].

The investigation of these modes is challenging from the experimental point of view because of their extreme confinement (and therefore techniques able to resolve momentum on the order of $\omega/v_F$ are required). A crucial point is also to address the temperature dependence of the optical properties, since *e-e* scattering, and therefore the transition from the collisionless to the hydrodynamic regime is very sensitive to the electronic temperature[19].

Finally, hydrodynamic phenomena occur in many materials beyond graphene[190] and electrons are not the only quasiparticles that display hydrodynamic behavior. This regime has also been predicted to happen for phonons[191,192] and magnons[193]. Future studies can address the optical response of more exotic quasiparticles when these enter the hydrodynamic regime. Combining high sample quality and advanced experimental techniques now enables exploration of electron-electron interactions in graphene and promote novel strategies for quantum sensing of collective excitations in 2D materials.

**Electrodynamics of superconductors probed with ultra-confined graphene plasmons**

In this section we discuss the excitation of collective modes in a superconductor (SC) by acoustic graphene plasmons (AGPs) and quantum emitters, that otherwise would be optically silent to far-field radiation. As an example, we present the coupling of the AGPs to the Higgs mode of a SC. A SC supports an abundance of collective modes defining its optical response[194]. These are:

- The Higgs mode[195,196], associated with fluctuations of the amplitude of the order-parameter, and appearing above the superconducting gap, $2\Delta$, where $\Delta$ is the amplitude of the superconducting order-parameter; it is, therefore, a gapped mode. However, since its energy range overlaps with the Bogoliubov quasiparticle continuum, the Higgs mode suffers from Landau damping, which contributes even further to the difficulty in observing this mode. Nevertheless, for energies close to $2\Delta$, the density of particle-hole excitations should be small enough so that the mode can be distinguished from the continuum.

- The Nambu-Goldstone mode[197] [22], associated with fluctuations of the phase of the order parameter, is in principle gapless, being the result of the spontaneous breaking of a continuous symmetry. The coupling of this mode to the electromagnetic field, however, results in a gap of the order of the plasmon energy[198].

- Collective charge excitations, dubbed intrinsic plasmons, which in a 2D SC disperse as the square root of the wave vector. In addition, in a layered superconductor a number of linear dispersive plasmon branches appear, due to the Coulomb interaction among charges in different layers. Also, and in particular, when Cooper pairs tunneling is allowed in a layered SC, Josephson plasmons (manifesting in the optical response along the direction perpendicular to the Cu-O planes, e. g., in cuprates SCs) also appear[199–202].

- Other modes are the Carlson-Goldman (CG) mode[203,204], which is a phase mode in charged SCs, and the Bardasis-Schrieffer (BS) mode[205–207], which is a gapped mode appearing in fluctuations in subdominant order-parameters in a s-wave superconductor, located at energies below the superconducting gap. The BS mode, similarly to the Higgs mode, does not couple linearly with far-field electromagnetic radiation. It has been proposed recently[208] that it can be made observable by inducing externally a supercurrent in a 2D SC placed in a microwave cavity. It has also been suggested that, due to the sub-gap nature of the BS mode, a (non-equilibrium) finite density of BS-polaritons can be achieved, producing a mixed *s+id* superconducting state.

In addition to all of these modes, a semi-infinite SC (or a thin SC film) also supports interfacial hybrid charge-radiation modes dubbed surface plasmon polaritons (SPPs)[209], first described by Keller[210,211], and which can be explored for extraordinary transmission through a perforated superconductor with subwavelength holes[212] and for low-loss plasmonics[213].

The energy scale for all these modes is given by the amplitude of the order parameter which lies in THz spectral range. Sources of THz radiation are, therefore, needed for probing the aforementioned modes. Although sources of THz radiation are nowadays widely available, not all of these modes couple to transverse far-field radiation. This is the case of the Higgs mode in a clean superconductor, the plasmon mode, and the CG and BS modes. On the other hand, these modes couple to longitudinal electromagnetic fields at finite wave vectors. Also, as it is well known, SPPs do not couple directly to far-field radiation due to kinematic reasons.

Electromagnetic sources of large wave-vectors can be found in the near field of scattered electromagnetic radiation, such as in surface plasmons polaritons and in quantum emitters decaying near an interface. Therefore, exciting the aforementioned modes in the SC requires sources generating near fields of THz radiation. An obvious choice of generating near fields is Scanning Near-field Optical Microscopy (SNOM)[214] when the tip of the microscope is illuminated by far-field THz radiation.

It is in the above context that (nonlocal) AGPs[57] emerge as a complementary near-field probe. When in close proximity to a metal or a superconductor, AGPs have their electrostatic interaction screened by the presence of the charges in the metal or the SC. The AGPs then disperse linearly with a speed slightly above the Fermi velocity of the charge carriers in graphene and, therefore, even for THz radiation, they carry very large wave vectors. Then, it is conceivable that AGPs can be used as a probe of the elusive collective excitations of a SC. In particular, we foresee the possibility to probe, beyond the collective modes, also the anisotropy of the order parameter[215], the Cooper pairs size as its built into the nonlocal conductivity, and probing the London penetration depth[216]. Since AGPs are ultra-confined modes, thus leading to spatially localized high intense fields, it is conceivable to explore the nonlinear response of a SC, leading to higher harmonics generation[217]. Naturally, from a theoretical perspective, the description of the interaction of the SC modes with AGPs requires a detailed calculation of the nonlocal optical conductivity of the SC (together with graphene's nonlocal optical conductivity[10,29,218,219]). The simplest and nontrivial description of the conductivity of a bulk SC using BCS theory, and including nonlocal corrections to leading-order, was given by Keller[210]. In a recent work[25], using Keller's conductivity, it was shown that AGPs can couple to the Higgs mode leading to an anti-crossing in the dispersion relation of the AGPs near the Higgs mode frequency $2\Delta/\hbar$ (cf. **Figure 5**).

The linear optical conductivity of a BCS-like superconductor in the weak nonlocal regime[210] shows a sharp feature at frequencies close to $2\Delta/\hbar$, which can be attributed to the excitation of the Higgs mode. For a SC-insulator interface, this feature leads to a stop-band in the SC's SPP-dispersion, analogous to the avoided crossing between the bulk plasmon polariton and the SPP conventional metal-insulator interfaces. In the SC-insulator case, however, the energy gap between the two ensuing branches is extremely narrow (a few μeV in the clean limit, for $2\Delta/\hbar$ ~28 meV). Moreover, unless the relaxation rate is unrealistically small, dissipative effects blur the distinction between the two branches, rendering this feature essentially unobservable. However, if a graphene sheet is deposited in close proximity to the SC surface, there is clearly hybridization between AGPs and the Higgs mode, as already noted. The features of this hybrid mode can be tuned by controlling graphene's doping, the graphene-SC distance or the system's temperature. To observe the hybrid mode, the mismatch between AGPs and far-field radiation wave vector has to be overcome. This can be achieved by patterning graphene into micro-ribbons, separated from the SC by a thin insulating layer (cf. **Figure 5**). Another possibility, experimentally feasible[57], is to nano-fabricate a metallic grid on graphene, with a few-layer hexagonal Boron-Nitride as spacer layer. Alternatively, one can study the decay rate of a quantum emitter in the proximity of the graphene-insulator-SC heterostructure. It has been shown recently[25] that a strong feature can be observed in the Purcell factor of such a configuration. This feature is reminiscent of a Fano resonance and suggests a competition between two decaying channels, one associated with the coherent hybrid mode and the other with the dissipative excitation of independent particle pairs. It should be noted that, in the vdW heterostructure considered in the discussion above, graphene is always separated from the superconductor by an insulating material (see Fig. 5, panel a). Since the SC order parameter decays exponentially with distance from the SC surface, this separation should be enough to hinder the induction of a superconducting gap in graphene by proximity effect.

An alternative method to excite Higgs modes has been proposed recently, in which a superconductor is driven out of equilibrium by short terahertz pulses, resulting in oscillations of the amplitude of the order parameter[220]. In the same vein, transient superconductivity induced by femtosecond mid-infrared pulses, besides activating the elusive Higgs mode, can be used to amplify terahertz radiation[221].

In conclusion, AGPs are emerging as a new experimental tool to unveil the subtle collective excitations in SCs, and can, in principle, be applied to both BCS-like SCs and to unconventional layered and 2D SCs. The latter possibility is particularly exciting as very little is known about 2D SCs. Since van der Waals heterostructures of 2D materials are routinely fabricated, the prospect of combing 2D SCs with other 2D materials, including graphene, becomes viable.

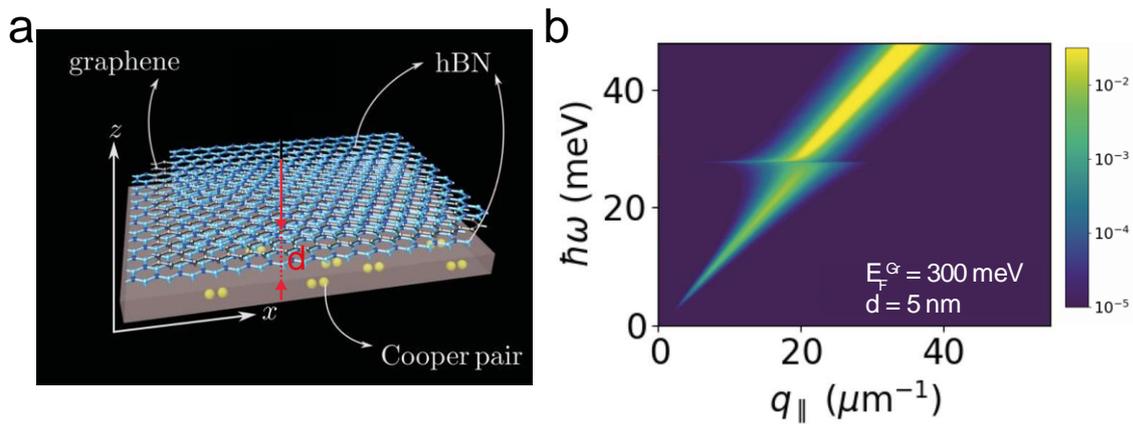

**Figure 5: Loss function of the heterostructure graphene-hBN-superconductor.** a: Sketch of the heterostructure, showing the superconductor material with cooper pairs, separated by a thin layer if hBN (thickness d) to graphene top layer. b: loss function calculated for Fermi energy of 0.4 eV and graphene-SC separation of 5 nm. The nonlocal conductivities of the SC and of graphene were used in the calculation. A level repulsion between the AGP dispersion and the Higgs dispersion is clearly seen.

**Plasmon-magnon coupling in magnetic van der Waals heterostructures**

The family of 2D materials beyond graphene is now very large and contains also many magnetic crystals[222]. Their electronic and magnetic properties can be studied via density functional theory (DFT), although great care needs to be exercised since the often-used DFT+U approaches fail in predicting their ground-state properties. Indeed, it has been recently shown[223] that short-range *e-e* interactions play a pivotal role, and these are well captured by hybrid functionals. When insulating magnetic materials are doped, either via the electric field effect or chemically, their spin waves are expected to hybridize with the charge collective modes (plasmons) of the itinerant electron system. Microscopically, strong spin-orbit coupling is at the origin of this hybridization. Spin-orbit coupling is also responsible for inducing topological behavior[224] and non-reciprocity[225] in magnons, giving rise to the intriguing possibility that the hybrid excitations inherit such features. Antiferromagnetic materials are expected to host spin waves in the THz spectral range, where plasmons are weakly damped provided that the electronic quality of the material is high. We therefore expect nanophotonic probes to be extremely useful in the near future to unveil this coupling, potentially enabling neutron-free spectroscopy of spin waves in mono- and bi-layer van der Waals magnets. We also envision the exciting possibility to study the coupling of graphene plasmons with spin waves in van der Waals heterostructures comprising graphene and magnetic materials, provided that they harbor substantial interfacial spin-orbit coupling.

**Topological plasmonics**

Plasmon modes can be sensitive to the intricate winding of electronic wavefunctions in a quantum material. A prime example can be found in topological materials which feature electrons that are characterized not only by its momentum and effective mass, but also by new band geometric quantities[226]. One prominent band geometric quantity is Berry curvature which produces a Hall effect in the absence of a magnetic field[226,227]. The plasmonic collective modes of bulk electrons in such anomalous Hall materials can similarly inherit chirality in the absence of a magnetic field. These chiral "Berry" plasmons become intrinsically squeezed along the edges of a sample[228] and possess distinct chiral dispersion relations for plasmon waves propagating in a clock-wise vs an anti-clockwise direction[228,229]; they can allow to achieve zero-field non-reciprocal plasmonic propagation[230]. Such plasmons in topological materials display how to go beyond the simple Landau Fermi liquid paradigm that characterizes the collective modes of metals[231,232], with plasmons that are sensitive to their unique electron wavefunction characteristics. This wavefunction dependent behavior is intrinsic (available even in unstructured quantum materials) and distinct from that obtained from extrinsic nano-structuring such as those found in spatially-patterned topological photonic systems[233,234].

Another example of wavefunction sensitivity can be found in the plasmonic dynamics of electrons in topological boundary modes (TBMs). Such TBMs include gapless edge spectra in two/three-dimensional topological insulators, or the "open-segment" Fermi-arc surface states at the surfaces of Weyl/Dirac semimetals[235]. The plasmonic collective modes of electrons in these TBMs are similarly morphed (**Figure 6**a) with the collective motion of carriers in the topological boundary mode fundamentally changing its plasmonic properties. For instance, plasmons in Fermi-arc surface states become hyperbolic[236,237]: they exhibit open iso-frequency contours that persist to large momenta that can enable extreme compression of light. This arises from the unusual electron dynamics close to the boundaries of topological materials[236–239].

The gapless TBMs in two-dimensional topological insulators also host gapless plasmonic modes that are localized along topological domain walls; owing to the suppressed back scattering in such modes as well as a frequency mismatch with bulk modes, these domain wall plasmons can have ultra-long life times[240]. Surprisingly, while predominantly determined by electron dynamics in the TBMs, TBM plasmons can also be sensitive to bulk electron characteristics[240], with their dispersion relations becoming stiffened or split by bulk Hall motion[228,236,239,240]. This sensitivity may be useful in tracking the unusual coupled bulk/edge electron dynamics in topological materials.

Plasmons by themselves can also possess a topological/geometrical structure. Owing to the helicity of light, surface plasmon polaritons at the surface of a 3D metal possess a transverse spin that becomes locked to its momentum[241]. In two-dimensional metals, magneto-plasmons (at a finite applied magnetic field) can possess a finite Chern number[242], with topological plasmonic edge modes that persist to infrared frequencies in a plasmonic crystal[243] (cf. **Figure 6**b). In the deep subwavelength limit, this geometry/topology arises from a "hidden" internal structure of plasmons[244] – longitudinal electric plasmons are not only characterized by their electric fields (that are easily imaged by optical techniques), but also by the spatial pattern of the dynamical current density of the electrons in the metal (typically "hidden" from conventional probes). The latter cants away from the electric field (and in-plane momentum) when there is a Hall effect forming a "pseudo-spin" structure for deep subwavelength plasmons that is ubiquitous in two-dimensional metals, see **Figure 6**c. Coupling between "Pseudo-spin" (current density) and orbital (momentum) degrees of freedom lead to unusual propagation of plasmons (e.g., plasmons that do not obey classical ray optics[244]), spawning a type of "spin-orbit" coupling for plasmons and new ways to control their trajectories, **Figure 6**d.

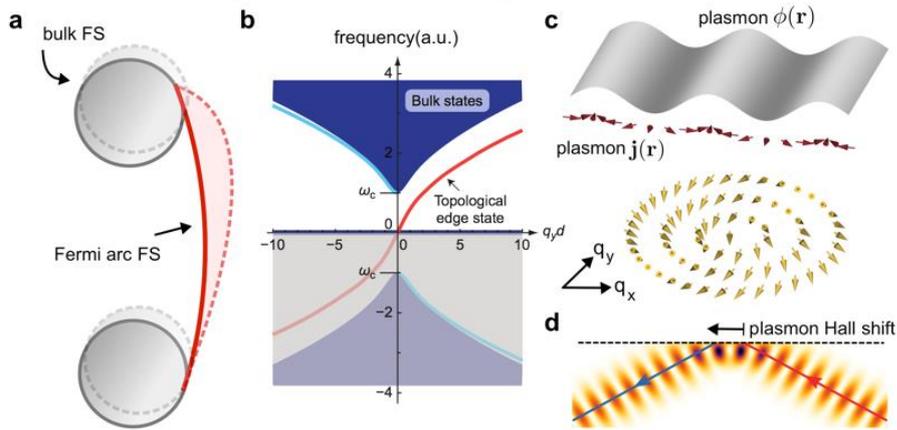

**Figure 6 : a** Carrier dynamics in topological boundary modes (TBMs) can yield unusual collective dynamics. For example, while the collective modes of bulk the Fermi surface (FS, gray) in a Weyl semimetal yield conventional bulk (and surface plasmons)[239], the unusual chiral motion of the Fermi arc Fermi surface (FS, red) yield hyperbolic surface plasmonic modes[236,237]. Solid lines denote equilibrium FS, dashed lines indicate deformed FS tracking the collective motion. **b** Plasmons can also possess their own topological structure. In two-dimensional magneto-plasmon modes can be characterized by a non-zero Chern number and possess a topologically protected chiral edge state, adapted from Ref[243]. **c** Plasmons can even possess an emergent "pseudo-spinor" degree of freedom that arises from the (top) local current density pattern (**j(r)**, red arrows) in a metal as a plasmon wave propagates; gray surface indicates electric potential ϕ(**r**) of the propagating plasmonic wave. (bottom) Oscillating current density direction is locked to momentum and can possess non-trivial hedgehog- like textures (yellow arrows), e.g., shown for a gapped bulk magnetoplasmon mode (bottom). When coupled with its momenta (e.g., by applying a magnetic field), this can produce unusual non-classical ray optics trajectories of plasmons **d** such as a real space non-reciprocal displacement (plasmon Hall shift) between incident (red) and reflected plasmon (blue) beams, adapted from Ref[244].

While electronic bandstructure is typically taken as a static/immutable quantity, large electric fields can be used to radically change the electronic bandstructure[245]. Such Floquet band engineering is typically achieved with ultra-strong incident fields (e.g., found in ultra-shot pulses[246]); in contrast, large oscillating electric fields associated with strongly confined plasmons in two-dimensional materials (such as graphene) can be readily obtained even with small amplitudes of incident light opening the way for a type of *plasmonic Floquet engineering*. Such engineering can be effected in a target material by proximal coupling with graphene plasmons (standard van der Waals stacking techniques enable easy integration).

Strikingly, when *plasmonic Floquet engineering* works in concert with "electron wavefunction sensitive" plasmons, a type of dynamical symmetry breaking can ensue. For instance, linearly polarized light irradiated on a graphene plasmonic disk can induce spontaneous symmetry breaking with plasmons rotating in either clockwise or anti-clockwise directions[247]. Such a collective mode phase displays a broken symmetry distinct from that of the ground (metallic) state and shows how plasmon motion can be decoupled from its host material. In such circumstances, the plasmon motion takes on its own separate life becoming a dynamical variable. While Ref[245] provides a graphene-based example, other systems, e.g., josephson plasmonics and superconducting metamaterials[199,209,248], can also exhibit strong plasmonic nonlinearities that may lead to new types of dynamical symmetry breaking. These raise the tantalizing prospect of the stabilization of out-of-equilibrium plasmonic phases in materials that by themselves (*i.e.* in their native equilibrium state) do not exhibit broken symmetry states.

**Moiré systems**

Recent experimental progress[164,249] has enabled precise control of the twist angle between the crystal axes of superimposed two-dimensional materials, leading to a new class of quantum materials dubbed "twisted van der Waals heterostructures". The most prominent example is twisted bilayer graphene[250]. In these structures moiré patterns form due to the combination of rotation angle and lattice constant mismatch between different layers, creating a superlattice that spatially modulates the tunneling amplitude for electrons between the layers[250,251]. This profoundly modifies the behavior of electrons in the heterostructure. The impact is particularly dramatic in twisted bilayer graphene where a plethora of new symmetry broken phases have been observed[164,249,252–254] and many more are theoretically predicted[255].

Despite extensive studies on twisted bilayer graphene, based on transport and scanning techniques, and several interesting theoretical predictions[256–260], a complete study of its optical properties[261,262], in particular at low temperatures, where the symmetry-broken states arise is still lacking. In this context, nanophotonics probes like s-SNOM[261,262] and spectroscopy are ideal candidates since they are able to conjugate high spatial resolution (down to $\approx 10$nm), with energies that are matching the expected optical features, often in the mid-Infrared or Terahertz (THz) frequency ranges (cf. **Figure 7**a-b). In particular, it has been recently predicted that optical properties can discriminate between different symmetry-broken states[260] and probe important details of the band structure of twisted bilayer graphene[258].

Moreover, the ability to combine different quantum materials (including superconductors, ferromagnets, etc) into twisted heterostructures and the interaction of their emergent quasiparticles with photons offers a virtually infinite amount of enticing physical phenomena to be explored.

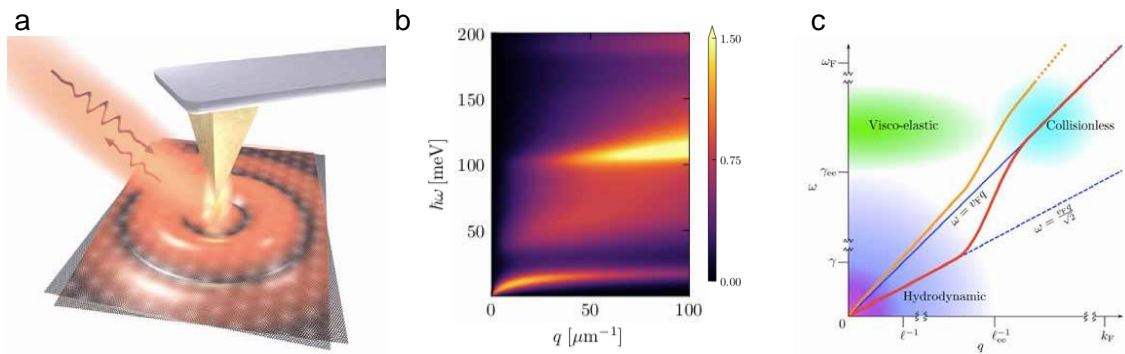

**Figure 7. a**: Illustration of a s-SNOM experiment on twisted bilayer graphene: an AFM metallic tip is illuminated by infrared light, which provides sufficient momentum to launch a collective excitation in twisted bilayer graphene. The plasmon can, in turn (e.g. by reflection from an edge or interface), scatter into light at the tip. This scattered light is detected by a photodetector. The curved arrows indicate the light impinging on the tip (coming from the laser) as well as the scattered light (going to the detector). Adapted from ref[262]. **b**: 2D plots of the energy loss function L(q, ω) at θ = 1.05°. Adapted from ref[258]. **c**: Sketch of the wavevector-frequency plane showing the relevant frequency and length scales, and the acoustic plasmon dispersion (red and orange lines) for two different values of the screening from an external metallic gate (orange line corresponds to moderate screening, red line to very strong screening). The blue solid line is the electron dispersion $\omega = v_F q$ while the blue dashed line is the sound dispersion $\omega = v_F q/\sqrt{2}$. Different regimes of linear response are highlighted. In the hydrodynamic regime (blue shaded region) the Navier-Stokes equation is applicable to the electron liquid. In the overdamped regime (magenta shaded region) the Navier-Stokes equation is still applicable but plasmons are strongly damped. In the visco-elastic regime (green shaded region) Navier-Stokes equation can still be applied by considering a frequency-dependent complex viscosity, equivalent to an elastic shear modulus. Adapted from ref[187].

**Conclusion**

In summary, we analyzed the developing field of 2D quantum nanophotonics, highlighting the potential for quantum technologies, polaritonic sensing, new regimes of extreme light–matter interactions and polariton engineering. The unique combination of appropriate tools to investigate light–matter interactions at the nano-scale (SNOM, EELS, EL-STM) and quantum materials built atom by atom in basically any configuration (moiré, van der Waals heterostructures, nano-patterning) is truly a game changer. Nano-optics has now also become a comprehensive tool for the study of 2D quantum materials, and touch upon emerging phenomena in correlated and topological materials. Finally, a realistic pathway for the 2D quantum nanophotonics field to deliver technologically relevant applications can be envisioned, as a range of 2D materials can be grown in a scalable fashion, and scalable hetero-stacking of 2D materials has been demonstrated on waferscale[263,264], including twisted graphene[265].

**Acknowledgements:**
We thank Matteo Ceccanti for the 3D images in figure 2a-b. F.H.L.K. acknowledges support from the Government of Spain (FIS2017-91599-EXP; Severo Ochoa CEX2019-000910-S), Fundació Cellex, Fundació Mir-Puig, and Generalitat de Catalunya (CERCA, AGAUR, SGR 1656). Furthermore, the research leading to these results has received funding from the European Union's Horizon 2020 under grant agreements no.785219 (Graphene flagship Core2), no. 881603 (Graphene flagship Core3) and 820378 (Quantum flagship). This work was also supported by the ERC TOPONANOP under grant agreement n° 726001. I.T. acknowledges funding from the Spanish Ministry of Science, Innovation and Universities (MCIU) and State Research Agency (AEI) via the Juan de la Cierva fellowship n. FJC2018-037098-I. F.H.L.K and A. R-P acknowledge BIST Ignite Programme grant from the Barcelona Institute of Science and Technology (QEE2DUP). N.M.R.P. acknowledges support from the European Commission through the project "Graphene-Driven Revolutions in ICT and Beyond" (Ref. No. 881603, CORE 3), COMPETE 2020, PORTUGAL 2020, FEDER and the Portuguese Foundation for Science and Technology (FCT) through project POCI-01-0145-FEDER-028114, and the Portuguese Foundation for Science and Technology (FCT) in the framework of the Strategic Financing UID/FIS/04650/2019. N. A. M. is a VILLUM Investigator supported by VILLUM FONDEN (Grant No. 16498) and Independent Research Fund Denmark (Grant No. 7026-00117B). The Center for Nano Optics is financially supported by the University of Southern Denmark (SDU 2020 funding). The Center for Nanostructured Graphene (CNG) is sponsored by the Danish National Research Foundation (Project No. DNRF103). J.C.W.S acknowledges support from the National Research Foundation (NRF) Singapore under its NRF fellowship programme award number NRF-NRFF2016-05 and the Ministry of Education (MOE) Singapore under its MOE AcRF Tier 3 Award MOE2018-T3-1-002.

**References:**
(1) Song, J. C. W.; Gabor, N. M. Electron Quantum Metamaterials in van Der Waals Heterostructures. *Nat. Nanotechnol.* **2018**, *13* (November).
(2) Geim, A. K.; Grigorieva, I. V. Van Der Waals Heterostructures. *Nature*. 2013.
(3) Low, T.; Chaves, A.; Caldwell, J. D.; Kumar, A.; Fang, N. X.; Avouris, P.; Heinz, T. F.; Guinea, F.; Martin-Moreno, L.; Koppens, F. Polaritons in Layered Two-Dimensional Materials. *Nature Materials*. 2017.
(4) Basov, D. N.; Fogler, M. M.; García De Abajo, F. J. Polaritons in van Der Waals Materials. *Science*. 2016.
(5) Grigorenko, A. N.; Polini, M.; Novoselov, K. S. Graphene Plasmonics. *Nature Photonics*. 2012.
(6) Rivera, N.; Christensen, T.; Narang, P. Phonon Polaritonics in Two-Dimensional Materials. *Nano Lett.* **2019**.
(7) Iranzo, D. A.; Nanot, S.; Dias, E. J. C.; Epstein, I.; Peng, C.; Efetov, D. K.; Lundeberg, M. B.; Parret, R.; Osmond, J.; Hong, J. Y.; et al. Probing the Ultimate Plasmon Confinement Limits with a van Der Waals Heterostructure. *Science (80-. ).* **2018**.
(8) Lundeberg, M. B.; Gao, Y.; Asgari, R.; Tan, C.; Duppen, B. Van; Autore, M.; Alonso-González, P.; Woessner, A.; Watanabe, K.; Taniguchi, T.; et al. Tuning Quantum Nonlocal Effects in Graphene Plasmonics. *Science (80-. ).* **2017**.
(9) Dias, E. J. C.; Iranzo, D. A.; Gonçalves, P. A. D.; Hajati, Y.; Bludov, Y. V.; Jauho, A.-P.; Mortensen, N. A.; Koppens, F. H. L.; Peres, N. M. R. Probing Nonlocal Effects in Metals with Graphene Plasmons. *Phys. Rev. B* **2018**, *97* (24), 245405.
(10) Gonçalves, P. A. D. *Plasmonics and Light–Matter Interactions in Two-Dimensional Materials and in Metal Nanostructures*; Springer Theses; Springer International Publishing: Cham, 2020.
(11) Gonçalves, P. A. D.; Christensen, T.; Rivera, N.; Jauho, A.-P.; Mortensen, N. A.; Soljačić, M. Plasmon–Emitter Interactions at the Nanoscale. *Nat. Commun.* **2020**, *11* (1), 366.
(12) Koppens, F. H. L.; Chang, D. E.; García De Abajo, F. J. Graphene Plasmonics: A Platform for

(13) Rivera, N.; Kaminer, I.; Zhen, B.; Joannopoulos, J. D.; Soljačić, M. Shrinking Light to Allow Forbidden Transitions on the Atomic Scale. *Science (80-. ).* **2016**.
(14) Rivera, N.; Kaminer, I. Light-Matter Interactions with Photonic Quasiparticles. **2020**.
(15) Gonçalves, P. A. D.; Stenger, N.; Cox, J. D.; Mortensen, N. A.; Xiao, S. Strong Light–Matter Interactions Enabled by Polaritons in Atomically Thin Materials. *Adv. Opt. Mater.* **2020**, *8* (5), 1901473.
(16) Schneider, C.; Glazov, M. M.; Korn, T.; Höfling, S.; Urbaszek, B. Two-Dimensional Semiconductors in the Regime of Strong Light-Matter Coupling. *Nat. Commun.* **2018**, *9* (1), 2695.
(17) Andersen, T. I.; Dwyer, B. L.; Sanchez-Yamagishi, J. D.; Rodriguez-Nieva, J. F.; Agarwal, K.; Watanabe, K.; Taniguchi, T.; Demler, E. A.; Kim, P.; Park, H.; et al. Electron-Phonon Instability in Graphene Revealed by Global and Local Noise Probes. *Science (80-. ).* **2019**.
(18) Kaminer, I.; Katan, Y. T.; Buljan, H.; Shen, Y.; Ilic, O.; López, J. J.; Wong, L. J.; Joannopoulos, J. D.; Soljačić, M. Efficient Plasmonic Emission by the Quantum Čerenkov Effect from Hot Carriers in Graphene. *Nat. Commun.* **2016**, *7* (1), ncomms11880.
(19) Bandurin, D. A.; Torre, I.; Kumar, R. K.; Ben Shalom, M.; Tomadin, A.; Principi, A.; Auton, G. H.; Khestanova, E.; Novoselov, K. S.; Grigorieva, I. V.; et al. Negative Local Resistance Caused by Viscous Electron Backflow in Graphene. *Science (80-. ).* **2016**, *351* (6277), 1055–1058.
(20) Polini, M.; Geim, A. K. Viscous Electron Fluids. *Phys. Today* **2020**, *73* (6), 28–34.
(21) Torre, I.; Tomadin, A.; Geim, A. K.; Polini, M. Nonlocal Transport and the Hydrodynamic Shear Viscosity in Graphene. *Phys. Rev. B* **2015**, *92* (16), 165433.
(22) Cox, J. D.; García De Abajo, F. J. Nonlinear Atom-Plasmon Interactions Enabled by Nanostructured Graphene. *Phys. Rev. Lett.* **2018**.
(23) Cox, J. D.; Silveiro, I.; Garciá De Abajo, F. J. Quantum Effects in the Nonlinear Response of Graphene Plasmons. *ACS Nano* **2016**.
(24) Fitzgerald, J. M.; Narang, P.; Craster, R. V.; Maier, S. A.; Giannini, V. Quantum Plasmonics. *Proc. IEEE* **2016**, *104* (12), 2307–2322.
(25) Costa, A. T.; Gonçalves, P. A. D.; Koppens, F. H. L.; Basov, D. N.; Mortensen, N. A.; Peres, N. M. R. Harnessing Ultra-Confined Graphene Plasmons to Probe the Electrodynamics of Superconductors. **2020**.
(26) Flick, J.; Rivera, N.; Narang, P. Strong Light-Matter Coupling in Quantum Chemistry and Quantum Photonics. *Nanophotonics* **2018**, *7* (9), 1479–1501.
(27) Yan, H.; Low, T.; Zhu, W.; Wu, Y.; Freitag, M.; Li, X.; Guinea, F.; Avouris, P.; Xia, F. Damping Pathways of Mid-Infrared Plasmons in Graphene Nanostructures. *Nat. Photonics* **2013**.
(28) Brar, V. W.; Jang, M. S.; Sherrott, M.; Lopez, J. J.; Atwater, H. A. Highly Confined Tunable Mid-Infrared Plasmonics in Graphene Nanoresonators. *Nano Lett.* **2013**.
(29) Gonçalves, P. A. D.; Peres, N. M. R. *An Introduction to Graphene Plasmonics*; 2016.
(30) Wang, G.; Chernikov, A.; Glazov, M. M.; Heinz, T. F.; Marie, X.; Amand, T.; Urbaszek, B. Colloquium: Excitons in Atomically Thin Transition Metal Dichalcogenides. *Rev. Mod. Phys.* **2018**.
(31) Scuri, G.; Zhou, Y.; High, A. A.; Wild, D. S.; Shu, C.; De Greve, K.; Jauregui, L. A.; Taniguchi, T.; Watanabe, K.; Kim, P.; et al. Large Excitonic Reflectivity of Monolayer MoSe2 Encapsulated in Hexagonal Boron Nitride. *Phys. Rev. Lett.* **2018**.
(32) Back, P.; Zeytinoglu, S.; Ijaz, A.; Kroner, M.; Imamoğlu, A. Realization of an Electrically Tunable Narrow-Bandwidth Atomically Thin Mirror Using Monolayer MoSe2. *Phys. Rev. Lett.* **2018**.
(33) Epstein, I.; Terrés, B.; Chaves, A. J.; Pusapati, V. V.; Rhodes, D. A.; Frank, B.; Zimmermann, V.; Qin, Y.; Watanabe, K.; Taniguchi, T.; et al. Near-Unity Light Absorption in a Monolayer Ws2 van Der Waals Heterostructure Cavity. *Nano Lett.* **2020**.
(34) Horng, J.; Martin, E. W.; Chou, Y.-H.; Courtade, E.; Chang, T.; Hsu, C.-Y.; Wentzel, M.-H.; Ruth, H. G.; Lu, T.; Cundiff, S. T.; et al. Perfect Absorption by an Atomically Thin Crystal. **2019**.
(35) Epstein, I.; Chaves, A. J. C.; Rhodes, D.; Frank, B.; Watanabe, K.; Taniguchi, T.; Giessen, H.;

Hone, J. C.; Peres, N.; Koppens, F. H. L. Highly Confined In-Plane Propagating Exciton-Polaritons on Monolayer Semiconductors. *2D Mater.* **2020**.

(36) Fang, H. H.; Han, B.; Robert, C.; Semina, M. A.; Lagarde, D.; Courtade, E.; Taniguchi, T.; Watanabe, K.; Amand, T.; Urbaszek, B.; et al. Control of the Exciton Radiative Lifetime in van Der Waals Heterostructures. *Phys. Rev. Lett.* **2019**.

(37) Horng, J.; Chou, Y.-H.; Chang, T.-C.; Hsu, C.-Y.; Lu, T.-C.; Deng, H. Engineering Radiative Coupling of Excitons in 2D Semiconductors. *Optica* **2019**.

(38) Rogers, C.; Gray, D.; Bogdanowicz, N.; Taniguchi, T.; Watanabe, K.; Mabuchi, H. Coherent Feedback Control of Two-Dimensional Excitons. *Phys. Rev. Res.* **2020**, *2* (1), 012029.

(39) Zeytinoğlu, S.; Roth, C.; Huber, S.; İmamoğlu, A. Atomically Thin Semiconductors as Nonlinear Mirrors. *Phys. Rev. A* **2017**, *96* (3), 031801.

(40) Wild, D. S.; Shahmoon, E.; Yelin, S. F.; Lukin, M. D. Quantum Nonlinear Optics in Atomically Thin Materials. *Phys. Rev. Lett.* **2018**, *121* (12), 123606.

(41) Ryou, A.; Rosser, D.; Saxena, A.; Fryett, T.; Majumdar, A. Strong Photon Antibunching in Weakly Nonlinear Two-Dimensional Exciton-Polaritons. *Phys. Rev. B* **2018**, *97* (23), 235307.

(42) Zeytinoğlu, S.; İmamoğlu, A. Interaction-Induced Photon Blockade Using an Atomically Thin Mirror Embedded in a Microcavity. *Phys. Rev. A* **2018**, *98* (5), 051801.

(43) Echarri, A. R.; Cox, J. D.; García de abajo, F. J. Quantum Effects in the Acoustic Plasmons of Atomically Thin Heterostructures: Publisher's Note. *Optica* **2019**.

(44) Cox, J. D.; García De Abajo, F. J. Single-Plasmon Thermo-Optical Switching in Graphene. *Nano Lett.* **2019**.

(45) Moreau, A.; Ciracì, C.; Mock, J. J.; Smith, D. R.; Hill, R. T.; Chilkoti, A.; Wang, Q.; Wiley, B. J. Controlled-Reflectance Surfaces with Film-Coupled Colloidal Nanoantennas. *Nature*. 2012.

(46) Baumberg, J. J.; Aizpurua, J.; Mikkelsen, M. H.; Smith, D. R. Extreme Nanophotonics from Ultrathin Metallic Gaps. *Nature Materials*. 2019.

(47) Huang, S.; Ming, T.; Lin, Y.; Ling, X.; Ruan, Q.; Palacios, T.; Wang, J.; Dresselhaus, M.; Kong, J. Ultrasmall Mode Volumes in Plasmonic Cavities of Nanoparticle-On-Mirror Structures. *Small* **2016**.

(48) Akselrod, G. M.; Argyropoulos, C.; Hoang, T. B.; Ciracì, C.; Fang, C.; Huang, J.; Smith, D. R.; Mikkelsen, M. H. Probing the Mechanisms of Large Purcell Enhancement in Plasmonic Nanoantennas. *Nat. Photonics* **2014**.

(49) Chikkaraddy, R.; Zheng, X.; Benz, F.; Brooks, L. J.; De Nijs, B.; Carnegie, C.; Kleemann, M. E.; Mertens, J.; Bowman, R. W.; Vandenbosch, G. A. E.; et al. How Ultranarrow Gap Symmetries Control Plasmonic Nanocavity Modes: From Cubes to Spheres in the Nanoparticle-on-Mirror. *ACS Photonics* **2017**.

(50) Hoang, T. B.; Akselrod, G. M.; Argyropoulos, C.; Huang, J.; Smith, D. R.; Mikkelsen, M. H. Ultrafast Spontaneous Emission Source Using Plasmonic Nanoantennas. *Nat. Commun.* **2015**.

(51) Feng, N. N.; Brongersma, M. L.; Dal Negro, L. Metal-Dielectric Slot-Waveguide Structures for the Propagation of Surface Plasmon Polaritons at 1.55 μ/M. *IEEE J. Quantum Electron.* **2007**.

(52) Khurgin, J. B. How to Deal with the Loss in Plasmonics and Metamaterials. *Nat. Nanotechnol.* **2015**.

(53) Ciracì, C.; Hill, R. T.; Mock, J. J.; Urzhumov, Y.; Fernández-Domínguez, A. I.; Maier, S. A.; Pendry, J. B.; Chilkoti, A.; Smith, D. R. Probing the Ultimate Limits of Plasmonic Enhancement. *Science (80-. ).* **2012**.

(54) Teperik, T. V.; Nordlander, P.; Aizpurua, J.; Borisov, A. G. Robust Subnanometric Plasmon Ruler by Rescaling of the Nonlocal Optical Response. *Phys. Rev. Lett.* **2013**.

(55) Toscano, G.; Raza, S.; Jauho, A.-P.; Mortensen, N. A.; Wubs, M. Modified Field Enhancement and Extinction by Plasmonic Nanowire Dimers Due to Nonlocal Response. *Opt. Express* **2012**.

(56) Kleemann, M. E.; Chikkaraddy, R.; Alexeev, E. M.; Kos, D.; Carnegie, C.; Deacon, W.; De Pury, A. C.; Große, C.; De Nijs, B.; Mertens, J.; et al. Strong-Coupling of WSe2 in Ultra-Compact Plasmonic Nanocavities at Room Temperature. *Nat. Commun.* **2017**.

(57) Iranzo, D. A.; Nanot, S.; Dias, E. J. C.; Epstein, I.; Peng, C.; Efetov, D. K.; Lundeberg, M. B.;

Parret, R.; Osmond, J.; Hong, J. Y.; et al. Probing the Ultimate Plasmon Confinement Limits with a van Der Waals Heterostructure. *Science (80-. ).* **2018**.

(58) Woessner, A.; Lundeberg, M. B.; Gao, Y.; Principi, A.; Alonso-González, P.; Carrega, M.; Watanabe, K.; Taniguchi, T.; Vignale, G.; Polini, M.; et al. Highly Confined Low-Loss Plasmons in Graphene–Boron Nitride Heterostructures. *Nat. Mater.* **2015**, *14* (4), 421–425.

(59) Epstein, I.; Alcaraz, D.; Huang, Z.; Pusapati, V.-V.; Hugonin, J.-P.; Kumar, A.; Deputy, X. M.; Khodkov, T.; Rappoport, T. G.; Hong, J.-Y.; et al. Far-Field Excitation of Single Graphene Plasmon Cavities with Ultracompressed Mode Volumes. *Science (80-. ).* **2020**, *368* (6496), 1219–1223.

(60) Dias, E. J. C.; Iranzo, D. A.; Gonçalves, P. A. D.; Hajati, Y.; Bludov, Y. V.; Jauho, A. P.; Mortensen, N. A.; Koppens, F. H. L.; Peres, N. M. R. Probing Nonlocal Effects in Metals with Graphene Plasmons. *Phys. Rev. B* **2018**.

(61) Kurman, Y.; Rivera, N.; Christensen, T.; Tsesses, S.; Orenstein, M.; Soljačić, M.; Joannopoulos, J. D.; Kaminer, I. Control of Semiconductor Emitter Frequency by Increasing Polariton Momenta. *Nat. Photonics* **2018**.

(62) Rodrigo, D.; Limaj, O.; Janner, D.; Etezadi, D.; Garcia de Abajo, F. J.; Pruneri, V.; Altug, H. Mid-Infrared Plasmonic Biosensing with Graphene. *Science (80-. ).* **2015**, *349* (6244), 165–168.

(63) Hu, H.; Yang, X.; Zhai, F.; Hu, D.; Liu, R.; Liu, K.; Sun, Z.; Dai, Q. Far-Field Nanoscale Infrared Spectroscopy of Vibrational Fingerprints of Molecules with Graphene Plasmons. *Nat. Commun.* **2016**, *7* (1), 12334.

(64) Autore, M.; Li, P.; Dolado, I.; Alfaro-Mozaz, F. J.; Esteban, R.; Atxabal, A.; Casanova, F.; Hueso, L. E.; Alonso-González, P.; Aizpurua, J.; et al. Boron Nitride Nanoresonators for Phonon-Enhanced Molecular Vibrational Spectroscopy at the Strong Coupling Limit. *Light Sci. Appl.* **2018**.

(65) Benz, F.; Schmidt, M. K.; Dreismann, A.; Chikkaraddy, R.; Zhang, Y.; Demetriadou, A.; Carnegie, C.; Ohadi, H.; De Nijs, B.; Esteban, R.; et al. Single-Molecule Optomechanics in "Picocavities." *Science (80-. ).* **2016**.

(66) Thomas, A.; Lethuillier-Karl, L.; Nagarajan, K.; Vergauwe, R. M. A.; George, J.; Chervy, T.; Shalabney, A.; Devaux, E.; Genet, C.; Moran, J.; et al. Tilting a Ground-State Reactivity Landscape by Vibrational Strong Coupling. *Science (80-. ).* **2019**.

(67) Tserkezis, C.; Fernández-Domínguez, A. I.; Gonçalves, P. A. D.; Todisco, F.; Cox, J. D.; Busch, K.; Stenger, N.; Bozhevolnyi, S. I.; Mortensen, N. A.; Wolff, C. On the Applicability of Quantum-Optical Concepts in Strong-Coupling Nanophotonics. *Reports Prog. Phys.* **2020**, *83* (8), 082401.

(68) Roques-Carmes, C.; Rivera, N.; Joannopoulos, J. D.; Soljačić, M.; Kaminer, I. Nonperturbative Quantum Electrodynamics in the Cherenkov Effect. *Phys. Rev. X* **2018**, *8* (4), 041013.

(69) Rivera, N.; Flick, J.; Narang, P. Variational Theory of Nonrelativistic Quantum Electrodynamics. *Phys. Rev. Lett.* **2019**, *122* (19), 193603.

(70) Rivera, N.; Wong, L. J.; Joannopoulos, J. D.; Soljačić, M.; Kaminer, I. Light Emission Based on Nanophotonic Vacuum Forces. *Nat. Phys.* **2019**, *15* (12), 1284–1289.

(71) Lundeberg, M. B.; Gao, Y.; Woessner, A.; Tan, C.; Alonso-González, P.; Watanabe, K.; Taniguchi, T.; Hone, J.; Hillenbrand, R.; Koppens, F. H. L. Thermoelectric Detection and Imaging of Propagating Graphene Plasmons. *Nat. Mater.* **2017**.

(72) Woessner, A.; Gao, Y.; Torre, I.; Lundeberg, M. B.; Tan, C.; Watanabe, K.; Taniguchi, T.; Hillenbrand, R.; Hone, J.; Polini, M.; et al. Electrical 2π Phase Control of Infrared Light in a 350-Nm Footprint Using Graphene Plasmons. *Nat. Photonics* **2017**, *11* (7), 421–424.

(73) De Vega, S.; García De Abajo, F. J. Plasmon Generation through Electron Tunneling in Graphene. *ACS Photonics* **2017**.

(74) Guerrero-Becerra, K. A.; Tomadin, A.; Polini, M. Electrical Plasmon Injection in Double-Layer Graphene Heterostructures. *Phys. Rev. B* **2019**.

(75) Cuadra, J.; Baranov, D. G.; Wersäll, M.; Verre, R.; Antosiewicz, T. J.; Shegai, T. Observation of Tunable Charged Exciton Polaritons in Hybrid Monolayer WS$_2$–Plasmonic Nanoantenna System. *Nano Lett.* **2018**, *18* (3), 1777–1785.

(76)  Liu, X.; Galfsky, T.; Sun, Z.; Xia, F.; Lin, E. C.; Lee, Y. H.; Kéna-Cohen, S.; Menon, V. M. Strong Light-Matter Coupling in Two-Dimensional Atomic Crystals. *Nat. Photonics* **2014**.

(77)  Krasnok, A.; Lepeshov, S.; Alú, A. Nanophotonics with 2D Transition Metal Dichalcogenides [Invited]. *Opt. Express* **2018**, *26* (12), 15972.

(78)  Khitrova, G.; Gibbs, H. M.; Kira, M.; Koch, S. W.; Scherer, A. Vacuum Rabi Splitting in Semiconductors. *Nat. Phys.* **2006**, *2* (2), 81–90.

(79)  Flatten, L. C.; He, Z.; Coles, D. M.; Trichet, A. A. P.; Powell, A. W.; Taylor, R. A.; Warner, J. H.; Smith, J. M. Room-Temperature Exciton-Polaritons with Two-Dimensional WS2. *Sci. Rep.* **2016**, *6* (1), 33134.

(80)  Dufferwiel, S.; Lyons, T. P.; Solnyshkov, D. D.; Trichet, A. A. P.; Catanzaro, A.; Withers, F.; Malpuech, G.; Smith, J. M.; Novoselov, K. S.; Skolnick, M. S.; et al. Valley Coherent Exciton-Polaritons in a Monolayer Semiconductor. *Nat. Commun.* **2018**, *9* (1), 4797.

(81)  Zhang, L.; Gogna, R.; Burg, W.; Tutuc, E.; Deng, H. Photonic-Crystal Exciton-Polaritons in Monolayer Semiconductors. *Nat. Commun.* **2018**, *9* (1), 713.

(82)  Chen, Y.; Miao, S.; Wang, T.; Zhong, D.; Saxena, A.; Chow, C.; Whitehead, J.; Gerace, D.; Xu, X.; Shi, S.-F.; et al. Metasurface Integrated Monolayer Exciton Polariton. *Nano Lett.* **2020**, *20* (7), 5292–5300.

(83)  Wen, J.; Wang, H.; Wang, W.; Deng, Z.; Zhuang, C.; Zhang, Y.; Liu, F.; She, J.; Chen, J.; Chen, H.; et al. Room-Temperature Strong Light–Matter Interaction with Active Control in Single Plasmonic Nanorod Coupled with Two-Dimensional Atomic Crystals. *Nano Lett.* **2017**, *17* (8), 4689–4697.

(84)  Kleemann, M.-E.; Chikkaraddy, R.; Alexeev, E. M.; Kos, D.; Carnegie, C.; Deacon, W.; de Pury, A. C.; Große, C.; de Nijs, B.; Mertens, J.; et al. Strong-Coupling of WSe2 in Ultra-Compact Plasmonic Nanocavities at Room Temperature. *Nat. Commun.* **2017**, *8* (1), 1296.

(85)  Han, X.; Wang, K.; Xing, X.; Wang, M.; Lu, P. Rabi Splitting in a Plasmonic Nanocavity Coupled to a WS 2 Monolayer at Room Temperature. *ACS Photonics* **2018**, *5* (10), 3970–3976.

(86)  Geisler, M.; Cui, X.; Wang, J.; Rindzevicius, T.; Gammelgaard, L.; Jessen, B. S.; Gonçalves, P. A. D.; Todisco, F.; Bøggild, P.; Boisen, A.; et al. Single-Crystalline Gold Nanodisks on WS 2 Mono- and Multilayers for Strong Coupling at Room Temperature. *ACS Photonics* **2019**, *6* (4), 994–1001.

(87)  Hou, S.; Tobing, L. Y. M.; Wang, X.; Xie, Z.; Yu, J.; Zhou, J.; Zhang, D.; Dang, C.; Coquet, P.; Tay, B. K.; et al. Manipulating Coherent Light–Matter Interaction: Continuous Transition between Strong Coupling and Weak Coupling in MoS 2 Monolayer Coupled with Plasmonic Nanocavities. *Adv. Opt. Mater.* **2019**, *7* (22), 1900857.

(88)  Stührenberg, M.; Munkhbat, B.; Baranov, D. G.; Cuadra, J.; Yankovich, A. B.; Antosiewicz, T. J.; Olsson, E.; Shegai, T. Strong Light–Matter Coupling between Plasmons in Individual Gold Bi-Pyramids and Excitons in Mono- and Multilayer WSe 2. *Nano Lett.* **2018**, *18* (9), 5938–5945.

(89)  Antosiewicz, T. J.; Apell, S. P.; Shegai, T. Plasmon–Exciton Interactions in a Core–Shell Geometry: From Enhanced Absorption to Strong Coupling. *ACS Photonics* **2014**, *1* (5), 454–463.

(90)  Leng, H.; Szychowski, B.; Daniel, M.-C.; Pelton, M. Strong Coupling and Induced Transparency at Room Temperature with Single Quantum Dots and Gap Plasmons. *Nat. Commun.* **2018**, *9* (1), 4012.

(91)  Pelton, M.; Storm, S. D.; Leng, H. Strong Coupling of Emitters to Single Plasmonic Nanoparticles: Exciton-Induced Transparency and Rabi Splitting. *Nanoscale* **2019**, *11* (31), 14540–14552.

(92)  Wersäll, M.; Cuadra, J.; Antosiewicz, T. J.; Balci, S.; Shegai, T. Observation of Mode Splitting in Photoluminescence of Individual Plasmonic Nanoparticles Strongly Coupled to Molecular Excitons. *Nano Lett.* **2017**, *17* (1), 551–558.

(93)  Vasa, P.; Wang, W.; Pomraenke, R.; Lammers, M.; Maiuri, M.; Manzoni, C.; Cerullo, G.; Lienau, C. Real-Time Observation of Ultrafast Rabi Oscillations between Excitons and Plasmons in Metal Nanostructures with J-Aggregates. *Nat. Photonics* **2013**, *7* (2), 128–132.

(94) Törmä, P.; Barnes, W. L. Strong Coupling between Surface Plasmon Polaritons and Emitters: A Review. *Reports Prog. Phys.* **2015**, *78* (1), 013901.

(95) Sáez-Blázquez, R.; Feist, J.; Fernández-Domínguez, A. I.; García-Vidal, F. J. Enhancing Photon Correlations through Plasmonic Strong Coupling. *Optica* **2017**, *4* (11), 1363.

(96) Ojambati, O. S.; Chikkaraddy, R.; Deacon, W. D.; Horton, M.; Kos, D.; Turek, V. A.; Keyser, U. F.; Baumberg, J. J. Quantum Electrodynamics at Room Temperature Coupling a Single Vibrating Molecule with a Plasmonic Nanocavity. *Nat. Commun.* **2019**, *10* (1), 1049.

(97) Loo, V.; Arnold, C.; Gazzano, O.; Lemaître, A.; Sagnes, I.; Krebs, O.; Voisin, P.; Senellart, P.; Lanco, L. Optical Nonlinearity for Few-Photon Pulses on a Quantum Dot-Pillar Cavity Device. *Phys. Rev. Lett.* **2012**, *109* (16), 166806.

(98) Volz, T.; Reinhard, A.; Winger, M.; Badolato, A.; Hennessy, K. J.; Hu, E. L.; Imamoğlu, A. Ultrafast All-Optical Switching by Single Photons. *Nat. Photonics* **2012**, *6* (9), 605–609.

(99) Faraon, A.; Fushman, I.; Englund, D.; Stoltz, N.; Petroff, P.; Vučković, J. Coherent Generation of Non-Classical Light on a Chip via Photon-Induced Tunnelling and Blockade. *Nat. Phys.* **2008**, *4* (11), 859–863.

(100) Ramezani, M.; Halpin, A.; Fernández-Domínguez, A. I.; Feist, J.; Rodriguez, S. R.-K.; Garcia-Vidal, F. J.; Gómez Rivas, J. Plasmon-Exciton-Polariton Lasing. *Optica* **2017**, *4* (1), 31.

(101) Arnardottir, K. B.; Moilanen, A. J.; Strashko, A.; Törmä, P.; Keeling, J. Multimode Organic Polariton Lasing. **2020**.

(102) Byrnes, T.; Kim, N. Y.; Yamamoto, Y. Exciton–Polariton Condensates. *Nat. Phys.* **2014**, *10* (11), 803–813.

(103) Väkeväinen, A. I.; Moilanen, A. J.; Nečada, M.; Hakala, T. K.; Daskalakis, K. S.; Törmä, P. Sub-Picosecond Thermalization Dynamics in Condensation of Strongly Coupled Lattice Plasmons. *Nat. Commun.* **2020**, *11* (1), 3139.

(104) Hutchison, J. A.; Schwartz, T.; Genet, C.; Devaux, E.; Ebbesen, T. W. Modifying Chemical Landscapes by Coupling to Vacuum Fields. *Angew. Chemie Int. Ed.* **2012**, *51* (7), 1592–1596.

(105) Feist, J.; Galego, J.; Garcia-Vidal, F. J. Polaritonic Chemistry with Organic Molecules. *ACS Photonics* **2018**, *5* (1), 205–216.

(106) Amo, A.; Liew, T. C. H.; Adrados, C.; Houdré, R.; Giacobino, E.; Kavokin, A. V.; Bramati, A. Exciton–Polariton Spin Switches. *Nat. Photonics* **2010**, *4* (6), 361–366.

(107) Chen, Y.-J.; Cain, J. D.; Stanev, T. K.; Dravid, V. P.; Stern, N. P. Valley-Polarized Exciton–Polaritons in a Monolayer Semiconductor. *Nat. Photonics* **2017**, *11* (7), 431–435.

(108) Sun, Z.; Gu, J.; Ghazaryan, A.; Shotan, Z.; Considine, C. R.; Dollar, M.; Chakraborty, B.; Liu, X.; Ghaemi, P.; Kéna-Cohen, S.; et al. Optical Control of Room-Temperature Valley Polaritons. *Nat. Photonics* **2017**, *11* (8), 491–496.

(109) Gómez-Santos, G.; Stauber, T. Fluorescence Quenching in Graphene: A Fundamental Ruler and Evidence for Transverse Plasmons. *Phys. Rev. B - Condens. Matter Mater. Phys.* **2011**.

(110) Muschik, C. A.; Moulieras, S.; Bachtold, A.; Koppens, F. H. L.; Lewenstein, M.; Chang, D. E. Harnessing Vacuum Forces for Quantum Sensing of Graphene Motion. *Phys. Rev. Lett.* **2014**.

(111) Gaudreau, L.; Tielrooij, K. J.; Prawiroatmodjo, G. E. D. K.; Osmond, J.; De Abajo, F. J. G.; Koppens, F. H. L. Universal Distance-Scaling of Nonradiative Energy Transfer to Graphene. *Nano Lett.* **2013**.

(112) Mazzamuto, G.; Tabani, A.; Pazzagli, S.; Rizvi, S.; Reserbat-Plantey, A.; Schädler, K.; Navickaite, G.; Gaudreau, L.; Cataliotti, F. S.; Koppens, F.; et al. Single-Molecule Study for a Graphene-Based Nano-Position Sensor. *New J. Phys.* **2014**, *16*.

(113) Federspiel, F.; Froehlicher, G.; Nasilowski, M.; Pedetti, S.; Mahmood, A.; Doudin, B.; Park, S.; Lee, J. O.; Halley, D.; Dubertret, B.; et al. Distance Dependence of the Energy Transfer Rate from a Single Semiconductor Nanostructure to Graphene. *Nano Lett.* **2015**.

(114) Tisler, J.; Oeckinghaus, T.; Stöhr, R. J.; Kolesov, R.; Reuter, R.; Reinhard, F.; Wrachtrup, J. Single Defect Center Scanning Near-Field Optical Microscopy on Graphene. *Nano Lett.* **2013**.

(115) Silva Neto, M. B.; Szilard, D.; Rosa, F. S. S.; Farina, C.; Pinheiro, F. A. Characterizing Critical Phenomena via the Purcell Effect. *Phys. Rev. B* **2017**.

(116) Karanikolas, V. D.; Marocico, C. A.; Eastham, P. R.; Bradley, A. L. Near-Field Relaxation of a Quantum Emitter to Two-Dimensional Semiconductors: Surface Dissipation and Exciton Polaritons. *Phys. Rev. B* **2016**.

(117) Reserbat-Plantey, A.; Schädler, K. G.; Gaudreau, L.; Navickaite, G.; Güttinger, J.; Chang, D.; Toninelli, C.; Bachtold, A.; Koppens, F. H. L. Electromechanical Control of Nitrogen-Vacancy Defect Emission Using Graphene NEMS. *Nat. Commun.* **2016**, *7*, 10218.

(118) Schädler, K. G.; Ciancico, C.; Pazzagli, S.; Lombardi, P.; Bachtold, A.; Toninelli, C.; Reserbat-Plantey, A.; Koppens, F. H. L. Electrical Control of Lifetime-Limited Quantum Emitters Using 2D Materials. *Nano Lett.* **2019**, acs.nanolett.9b00916.

(119) Tielrooij, K. J.; Orona, L.; Ferrier, A.; Badioli, M.; Navickaite, G.; Coop, S.; Nanot, S.; Kalinic, B.; Cesca, T.; Gaudreau, L.; et al. Electrical Control of Optical Emitter Relaxation Pathways Enabled by Graphene. *Nat. Phys.* **2015**.

(120) Cano, D.; Ferrier, A.; Soundarapandian, K.; Reserbat-Plantey, A.; Scarafagio, M.; Tallaire, A.; Seyeux, A.; Marcus, P.; Riedmatten, H. de; Goldner, P.; et al. Fast Electrical Modulation of Strong Near-Field Interactions between Erbium Emitters and Graphene. *Nat. Commun.* **2020**, *11* (1), 4094.

(121) Zhang, Y. X.; Zhang, Y.; Mølmer, K. Surface Plasmon Launching by Polariton Superradiance. *ACS Photonics* **2019**.

(122) Manjavacas, A.; Thongrattanasiri, S.; Chang, D. E.; García De Abajo, F. J. Temporal Quantum Control with Graphene. *New J. Phys.* **2012**.

(123) Tetienne, J. P.; Dontschuk, N.; Broadway, D. A.; Stacey, A.; Simpson, D. A.; Hollenberg, L. C. L. Quantum Imaging of Current Flow in Graphene. *Sci. Adv.* **2017**.

(124) Tonndorf, P.; Schmidt, R.; Schneider, R.; Kern, J.; Buscema, M.; Steele, G. A.; Castellanos-Gomez, A.; van der Zant, H. S. J.; Michaelis de Vasconcellos, S.; Bratschitsch, R. Single-Photon Emission from Localized Excitons in an Atomically Thin Semiconductor. *Optica* **2015**.

(125) He, Y. M.; Clark, G.; Schaibley, J. R.; He, Y.; Chen, M. C.; Wei, Y. J.; Ding, X.; Zhang, Q.; Yao, W.; Xu, X.; et al. Single Quantum Emitters in Monolayer Semiconductors. *Nat. Nanotechnol.* **2015**.

(126) Srivastava, A.; Sidler, M.; Allain, A. V.; Lembke, D. S.; Kis, A.; Imamoglu, A. Optically Active Quantum Dots in Monolayer WSe 2. *Nat. Nanotechnol.* **2015**.

(127) Koperski, M.; Nogajewski, K.; Arora, A.; Cherkez, V.; Mallet, P.; Veuillen, J. Y.; Marcus, J.; Kossacki, P.; Potemski, M. Single Photon Emitters in Exfoliated WSe2 Structures. *Nat. Nanotechnol.* **2015**.

(128) Chakraborty, C.; Kinnischtzke, L.; Goodfellow, K. M.; Beams, R.; Vamivakas, A. N. Voltage-Controlled Quantum Light from an Atomically Thin Semiconductor. *Nat. Nanotechnol.* **2015**.

(129) Branny, A.; Wang, G.; Kumar, S.; Robert, C.; Lassagne, B.; Marie, X.; Gerardot, B. D.; Urbaszek, B. Discrete Quantum Dot like Emitters in Monolayer MoSe2: Spatial Mapping, Magneto-Optics, and Charge Tuning. *Appl. Phys. Lett.* **2016**.

(130) Tran, T. T.; Bray, K.; Ford, M. J.; Toth, M.; Aharonovich, I. Quantum Emission from Hexagonal Boron Nitride Monolayers. *Nat. Nanotechnol.* **2016**.

(131) Palacios-Berraquero, C.; Barbone, M.; Kara, D. M.; Chen, X.; Goykhman, I.; Yoon, D.; Ott, A. K.; Beitner, J.; Watanabe, K.; Taniguchi, T.; et al. Atomically Thin Quantum Light-Emitting Diodes. *Nat. Commun.* **2016**.

(132) Chakraborty, C.; Goodfellow, K. M.; Dhara, S.; Yoshimura, A.; Meunier, V.; Vamivakas, A. N. Quantum-Confined Stark Effect of Individual Defects in a van Der Waals Heterostructure. *Nano Lett.* **2017**.

(133) Peyskens, F.; Chakraborty, C.; Muneeb, M.; Van Thourhout, D.; Englund, D. Integration of Single Photon Emitters in 2D Layered Materials with a Silicon Nitride Photonic Chip. *Nat. Commun.* **2019**.

(134) Kumar, S.; Brotóns-Gisbert, M.; Al-Khuzheyri, R.; Branny, A.; Ballesteros-Garcia, G.; Sánchez-Royo, J. F.; Gerardot, B. D. Resonant Laser Spectroscopy of Localized Excitons in Monolayer WSe_2. *Optica* **2016**.

(135) Tran, T. T.; Kianinia, M.; Nguyen, M.; Kim, S.; Xu, Z. Q.; Kubanek, A.; Toth, M.; Aharonovich, I.

Resonant Excitation of Quantum Emitters in Hexagonal Boron Nitride. *ACS Photonics* **2018**.

(136) Palacios-Berraquero, C.; Kara, D. M.; Montblanch, A. R. P.; Barbone, M.; Latawiec, P.; Yoon, D.; Ott, A. K.; Loncar, M.; Ferrari, A. C.; Atatüre, M. Large-Scale Quantum-Emitter Arrays in Atomically Thin Semiconductors. *Nat. Commun.* **2017**, *8* (May), 1–6.

(137) Kern, J.; Niehues, I.; Tonndorf, P.; Schmidt, R.; Wigger, D.; Schneider, R.; Stiehm, T.; Michaelis de Vasconcellos, S.; Reiter, D. E.; Kuhn, T.; et al. Nanoscale Positioning of Single-Photon Emitters in Atomically Thin WSe2. *Adv. Mater.* **2016**.

(138) Klein, J.; Lorke, M.; Florian, M.; Sigger, F.; Sigl, L.; Rey, S.; Wierzbowski, J.; Cerne, J.; Müller, K.; Mitterreiter, E.; et al. Site-Selectively Generated Photon Emitters in Monolayer MoS2 via Local Helium Ion Irradiation. *Nat. Commun.* **2019**.

(139) Klein, J.; Sigl, L.; Gyger, S.; Barthelmi, K.; Florian, M.; Rey, S.; Taniguchi, T.; Watanabe, K.; Jahnke, F.; Kastl, C.; et al. Scalable Single-Photon Sources in Atomically Thin MoS2. **2020**.

(140) Mitterreiter, E.; Schuler, B.; Cochrane, K. A.; Wurstbauer, U.; Weber-Bargioni, A.; Kastl, C.; Holleitner, A. W. Atomistic Positioning of Defects in Helium Ion Treated Single-Layer MoS2. *Nano Lett.* **2020**.

(141) Rosławska, A.; Leon, C. C.; Grewal, A.; Merino, P.; Merino, P.; Merino, P.; Kuhnke, K.; Kern, K.; Kern, K. Atomic-Scale Dynamics Probed by Photon Correlations. *ACS Nano*. 2020.

(142) Yu, H.; Liu, G. Bin; Tang, J.; Xu, X.; Yao, W. Moiré Excitons: From Programmable Quantum Emitter Arrays to Spin-Orbit–Coupled Artificial Lattices. *Sci. Adv.* **2017**.

(143) Rivera, P.; Schaibley, J. R.; Jones, A. M.; Ross, J. S.; Wu, S.; Aivazian, G.; Klement, P.; Seyler, K.; Clark, G.; Ghimire, N. J.; et al. Observation of Long-Lived Interlayer Excitons in Monolayer MoSe2–WSe2 Heterostructures. *Nat. Commun.* **2015**, *6* (1), 6242.

(144) Montblanch, A. R.-P.; Kara, D. M.; Paradisanos, I.; Purser, C. M.; Feuer, M. S. G.; Alexeev, E. M.; Stefan, L.; Qin, Y.; Blei, M.; Wang, G.; et al. Confinement of Long-Lived Interlayer Excitons in WS$_2$/WSe$_2$ Heterostructures. **2020**.

(145) Zhang, N.; Surrente, A.; Baranowski, M.; Maude, D. K.; Gant, P.; Castellanos-gomez, A.; Plochocka, P. Moiré Intralayer Excitons in a MoSe 2 /MoS 2 Heterostructure. **2018**.

(146) Seyler, K. L.; Rivera, P.; Yu, H.; Wilson, N. P.; Ray, E. L.; Mandrus, D. G.; Yan, J.; Yao, W.; Xu, X. Signatures of Moiré-Trapped Valley Excitons in MoSe 2 /WSe 2 Heterobilayers. *Nature*. 2019.

(147) Jin, C.; Regan, E. C.; Yan, A.; Iqbal Bakti Utama, M.; Wang, D.; Zhao, S.; Qin, Y.; Yang, S.; Zheng, Z.; Shi, S.; et al. Observation of Moiré Excitons in WSe2/WS2 Heterostructure Superlattices. *Nature*. 2019.

(148) Alexeev, E. M.; Ruiz-Tijerina, D. A.; Danovich, M.; Hamer, M. J.; Terry, D. J.; Nayak, P. K.; Ahn, S.; Pak, S.; Lee, J.; Sohn, J. I.; et al. Resonantly Hybridized Excitons in Moiré Superlattices in van Der Waals Heterostructures. *Nature* **2019**, *567* (7746), 81–86.

(149) Delhomme, A.; Vaclavkova, D.; Slobodeniuk, A.; Orlita, M.; Potemski, M.; Basko, D.; Watanabe, K.; Taniguchi, T.; Mauro, D.; Barreteau, C.; et al. Flipping Exciton Angular Momentum with Chiral Phonons in MoSe 2 /WSe 2 Heterobilayers. *2D Mater.* **2020**.

(150) Baek, H.; Brotons-Gisbert, M.; Koong, Z. X.; Campbell, A.; Rambach, M.; Watanabe, K.; Taniguchi, T.; Gerardot, B. D. Highly Energy-Tunable Quantum Light from Moiré-Trapped Excitons. *Sci. Adv.* **2020**, *6* (37), eaba8526.

(151) Sung, J.; Zhou, Y.; Scuri, G.; Zólyomi, V.; Andersen, T. I.; Yoo, H.; Wild, D. S.; Joe, A. Y.; Gelly, R. J.; Heo, H.; et al. Broken Mirror Symmetry in Excitonic Response of Reconstructed Domains in Twisted MoSe2/MoSe2 Bilayers. *Nat. Nanotechnol.* **2020**, *15* (9), 750–754.

(152) Lundeberg, M. B.; Gao, Y.; Asgari, R.; Tan, C.; Van Duppen, B.; Autore, M.; Alonso-González, P.; Woessner, A.; Watanabe, K.; Taniguchi, T.; et al. Tuning Quantum Nonlocal Effects in Graphene Plasmonics. *Science (80-. ).* **2017**, *357* (6347), 187–191.

(153) Schuler, B.; Cochrane, K. A.; Kastl, C.; Barnard, E.; Wong, E.; Borys, N.; Schwartzberg, A. M.; Ogletree, D. F.; de Abajo, F. J. G.; Weber-Bargioni, A. Electrically Driven Photon Emission from Individual Atomic Defects in Monolayer WS2. **2019**.

(154) Alonso-González, P.; Nikitin, A. Y.; Gao, Y.; Woessner, A.; Lundeberg, M. B.; Principi, A.; Forcellini, N.; Yan, W.; Vélez, S.; Huber, A. J.; et al. Acoustic Terahertz Graphene Plasmons

Revealed by Photocurrent Nanoscopy. *Nat. Nanotechnol.* **2017**, *12* (1), 31–35.

(155) Alcaraz Iranzo, D.; Nanot, S.; Dias, E. J. C.; Epstein, I.; Peng, C.; Efetov, D. K.; Lundeberg, M. B.; Parret, R.; Osmond, J.; Hong, J.-Y.; et al. Probing the Ultimate Plasmon Confinement Limits with a van Der Waals Heterostructure. *Science (80-. ).* **2018**, *360* (6386), 291–295.

(156) Giuliani, G.; Vignale, G. *Quantum Theory of the Electron Liquid*; Cambridge University Press, 2005.

(157) Principi, A.; van Loon, E.; Polini, M.; Katsnelson, M. I. Confining Graphene Plasmons to the Ultimate Limit. *Phys. Rev. B* **2018**, *98* (3), 035427.

(158) Castro Neto, A. H.; Guinea, F.; Peres, N. M. R.; Novoselov, K. S.; Geim, A. K. The Electronic Properties of Graphene. *Rev. Mod. Phys.* **2009**, *81* (1), 109–162.

(159) Tang, H.-K.; Leaw, J. N.; Rodrigues, J. N. B.; Herbut, I. F.; Sengupta, P.; Assaad, F. F.; Adam, S. The Role of Electron-Electron Interactions in Two-Dimensional Dirac Fermions. *Science (80-. ).* **2018**, *361* (6402), 570–574.

(160) FEIBELMAN, P. Surface Electromagnetic Fields. *Prog. Surf. Sci.* **1982**, *12* (4), 287–407.

(161) LIEBSCH, A. ELECTRONIC EXCITATIONS AT METAL SURFACES. In *Photonic Probes of Surfaces*; Elsevier, 1995; pp 479–532.

(162) Persson, B. N. J.; Apell, P. Sum Rules for Surface Response Functions with Application to the van Der Waals Interaction between an Atom and a Metal. *Phys. Rev. B* **1983**, *27* (10), 6058–6065.

(163) P. A. D. Gonç̧alves, T. Christensen, N. M. R. Peres, A.-P. Jauho, I. Epstein, F. H. L. Koppens, M. Soljačić, and N. A. M. No Title. *Priv. Commun.*

(164) Cao, Y.; Fatemi, V.; Demir, A.; Fang, S.; Tomarken, S. L.; Luo, J. Y.; Sanchez-Yamagishi, J. D.; Watanabe, K.; Taniguchi, T.; Kaxiras, E.; et al. Correlated Insulator Behaviour at Half-Filling in Magic-Angle Graphene Superlattices. *Nature* **2018**.

(165) Lu, X.; Stepanov, P.; Yang, W.; Xie, M.; Aamir, M. A.; Das, I.; Urgell, C.; Watanabe, K.; Taniguchi, T.; Zhang, G.; et al. Superconductors, Orbital Magnets and Correlated States in Magic-Angle Bilayer Graphene. *Nature* **2019**, *574* (7780), 653–657.

(166) Polshyn, H.; Yankowitz, M.; Chen, S.; Zhang, Y.; Watanabe, K.; Taniguchi, T.; Dean, C. R.; Young, A. F. Large Linear-in-Temperature Resistivity in Twisted Bilayer Graphene. *Nat. Phys.* **2019**, *15* (10), 1011–1016.

(167) Haldane, F. D. M. Geometrical Description of the Fractional Quantum Hall Effect. *Phys. Rev. Lett.* **2011**.

(168) Girvin, S. M.; MacDonald, A. H.; Platzman, P. M. Magneto-Roton Theory of Collective Excitations in the Fractional Quantum Hall Effect. *Phys. Rev. B* **1986**.

(169) Kukushkin, I. V.; Smet, J. H.; Scarola, V. W.; Umansky, V.; Von Klitzing, K. Dispersion of the Excitations of Fractional Quantum Hall States. *Science (80-. ).* **2009**.

(170) Liou, S. F.; Haldane, F. D. M.; Yang, K.; Rezayi, E. H. Chiral Gravitons in Fractional Quantum Hall Liquids. *Phys. Rev. Lett.* **2019**.

(171) Yang, K. Acoustic Wave Absorption as a Probe of Dynamical Geometrical Response of Fractional Quantum Hall Liquids. *Phys. Rev. B* **2016**.

(172) Gurzhi, R. N.; Kopeliovich, A. I. Low-Temperature Electrical Conductivity of Pure Metals. *Sov. Phys. Uspekhi* **1981**, *24* (1), 17–41.

(173) Dyakonov, M. I.; Shur, M. S. Choking of Electron Flow: A Mechanism of Current Saturation in Field-Effect Transistors. *Phys. Rev. B* **1995**.

(174) Crossno, J.; Shi, J. K.; Wang, K.; Liu, X.; Harzheim, A.; Lucas, A.; Sachdev, S.; Kim, P.; Taniguchi, T.; Watanabe, K.; et al. Observation of the Dirac Fluid and the Breakdown of the Wiedemann-Franz Law in Graphene. *Science (80-. ).* **2016**, *351* (6277), 1058–1061.

(175) Ku, M. J. H.; Zhou, T. X.; Li, Q.; Shin, Y. J.; Shi, J. K.; Burch, C.; Anderson, L. E.; Pierce, A. T.; Xie, Y.; Hamo, A.; et al. Imaging Viscous Flow of the Dirac Fluid in Graphene. *Nature* **2020**, *583* (7817), 537–541.

(176) Moll, P. J. W.; Kushwaha, P.; Nandi, N.; Schmidt, B.; Mackenzie, A. P. Evidence for Hydrodynamic Electron Flow in PdCoO2. *Science (80-. ).* **2016**.

(177) Gooth, J.; Menges, F.; Kumar, N.; Süβ, V.; Shekhar, C.; Sun, Y.; Drechsler, U.; Zierold, R.; Felser, C.; Gotsmann, B. Thermal and Electrical Signatures of a Hydrodynamic Electron Fluid in Tungsten Diphosphide. *Nat. Commun.* **2018**, *9* (1), 4093.

(178) Vool, U.; Hamo, A.; Varnavides, G.; Wang, Y.; Zhou, T. X.; Kumar, N.; Dovzhenko, Y.; Qiu, Z.; Garcia, C. A. C.; Pierce, A. T.; et al. Imaging Phonon-Mediated Hydrodynamic Flow in WTe2 with Cryogenic Quantum Magnetometry. **2020**.

(179) Principi, A.; Vignale, G.; Carrega, M.; Polini, M. Bulk and Shear Viscosities of the Two-Dimensional Electron Liquid in a Doped Graphene Sheet. *Phys. Rev. B* **2016**, *93* (12), 125410.

(180) Coulter, J.; Sundararaman, R.; Narang, P. Microscopic Origins of Hydrodynamic Transport in the Type-II Weyl Semimetal $WP_2$. *Phys. Rev. B* **2018**, *98* (11), 115130.

(181) Torre, I.; Tomadin, A.; Geim, A. K.; Polini, M. Nonlocal Transport and the Hydrodynamic Shear Viscosity in Graphene. *Phys. Rev. B - Condens. Matter Mater. Phys.* **2015**.

(182) Raza, Sø.; Bozhevolnyi, S. I.; Wubs, M.; Asger Mortensen, N. Nonlocal Optical Response in Metallic Nanostructures. *Journal of Physics Condensed Matter*. 2015.

(183) Gao, H.; Dong, Z.; Levitov, L. Plasmonic Drag in a Flowing Fermi Liquid. **2019**.

(184) Svintsov, D. Emission of Plasmons by Drifting Dirac Electrons: A Hallmark of Hydrodynamic Transport. *Phys. Rev. B* **2019**.

(185) Svintsov, D. Hydrodynamic-to-Ballistic Crossover in Dirac Materials. *Phys. Rev. B* **2018**.

(186) Lucas, A.; Das Sarma, S. Electronic Sound Modes and Plasmons in Hydrodynamic Two-Dimensional Metals. *Phys. Rev. B* **2018**.

(187) Torre, I.; Vieira De Castro, L.; Van Duppen, B.; Barcons Ruiz, D.; Peeters, F. M.; Koppens, F. H. L.; Polini, M. Acoustic Plasmons at the Crossover between the Collisionless and Hydrodynamic Regimes in Two-Dimensional Electron Liquids. *Phys. Rev. B* **2019**.

(188) Sun, Z.; Basov, D. N.; Fogler, M. M. Universal Linear and Nonlinear Electrodynamics of a Dirac Fluid. *Proc. Natl. Acad. Sci. U. S. A.* **2018**.

(189) Phan, T. V.; Song, J. C. W.; Levitov, L. S. Ballistic Heat Transfer and Energy Waves in an Electron System. **2013**.

(190) Varnavides, G.; Jermyn, A. S.; Anikeeva, P.; Felser, C.; Narang, P. Electron Hydrodynamics in Anisotropic Materials. *Nat. Commun.* **2020**, *11* (1), 4710.

(191) Cepellotti, A.; Fugallo, G.; Paulatto, L.; Lazzeri, M.; Mauri, F.; Marzari, N. Phonon Hydrodynamics in Two-Dimensional Materials. *Nat. Commun.* **2015**, *6*, 1–7.

(192) Machida, Y.; Matsumoto, N.; Isono, T.; Behnia, K. Phonon Hydrodynamics and Ultrahigh–Room-Temperature Thermal Conductivity in Thin Graphite. *Science (80-. ).* **2020**, *367* (6475), 309–312.

(193) Ulloa, C.; Tomadin, A.; Shan, J.; Polini, M.; Van Wees, B. J.; Duine, R. A. Nonlocal Spin Transport as a Probe of Viscous Magnon Fluids. *Phys. Rev. Lett.* **2019**.

(194) Sun, Z.; Fogler, M. M.; Basov, D. N.; Millis, A. J. Collective Modes and Terahertz Near-Field Response of Superconductors. *arxiv* **2020**, *2001.03704*.

(195) Pekker, D.; Varma, C. M. Amplitude/Higgs Modes in Condensed Matter Physics. *Annu. Rev. Condens. Matter Phys.* **2015**.

(196) Shimano, R.; Tsuji, N. Higgs Mode in Superconductors. *Annu. Rev. Condens. Matter Phys.* **2020**.

(197) Kadowaki, K.; Kakeya, I. Longitudinal Josephson-Plasma Excitation in Direct Observation of the Nambu-Goldstone Mode in a Superconductor. *Phys. Rev. B - Condens. Matter Mater. Phys.* **1997**.

(198) Anderson, P. W. Plasmons, Gauge Invariance, and Mass. *Phys. Rev.* **1963**, *130* (1), 439–442.

(199) Laplace, Y.; Cavalleri, A. Josephson Plasmonics in Layered Superconductors. *Advances in Physics: X*. 2016.

(200) Dienst, A.; Casandruc, E.; Fausti, D.; Zhang, L.; Eckstein, M.; Hoffmann, M.; Khanna, V.; Dean, N.; Gensch, M.; Winnerl, S.; et al. Optical Excitation of Josephson Plasma Solitons in a Cuprate Superconductor. *Nat. Mater.* **2013**.

(201) Rajasekaran, S.; Casandruc, E.; Laplace, Y.; Nicoletti, D.; Gu, G. D.; Clark, S. R.; Jaksch, D.; Cavalleri, A. Parametric Amplification of a Superconducting Plasma Wave. *Nat. Phys.* **2016**, *12* (11), 1012–1016.

(202) Basov, D. N.; Timusk, T. Electrodynamics of High-$T_c$ Superconductors. *Rev. Mod. Phys.* **2005**, *77* (2), 721–779.

(203) Sharapov, S. G.; Beck, H. Effective Action Approach and Carlson-Goldman Mode in d-Wave Superconductors. *Phys. Rev. B - Condens. Matter Mater. Phys.* **2002**.

(204) Ohashi, Y.; Takada, S. Goldstone Mode in Charged Superconductivity: Theoretical Studies of the Carlson-Goldman Mode and Effects of the Landau Damping in the Superconducting State. *J. Phys. Soc. Japan* **1997**.

(205) Bardasis, A.; Schrieffer, J. R. Excitons and Plasmons in Superconductors. *Phys. Rev.* **1961**.

(206) Maiti, S.; Hirschfeld, P. J. Collective Modes in Superconductors with Competing $s$- and $d$-Wave Interactions. *Phys. Rev. B* **2015**, *92* (9), 094506.

(207) Maiti, S.; Maier, T. A.; Böhm, T.; Hackl, R.; Hirschfeld, P. J. Probing the Pairing Interaction and Multiple Bardasis-Schrieffer Modes Using Raman Spectroscopy. *Phys. Rev. Lett.* **2016**.

(208) Allocca, A. A.; Raines, Z. M.; Curtis, J. B.; Galitski, V. M. Cavity Superconductor-Polaritons. *Phys. Rev. B* **2019**, *99* (2), 020504.

(209) Stockman, M. I.; Kneipp, K.; Bozhevolnyi, S. I.; Saha, S.; Dutta, A.; Ndukaife, J.; Kinsey, N.; Reddy, H.; Guler, U.; Shalaev, V. M.; et al. Roadmap on Plasmonics. *J. Opt. (United Kingdom)* **2018**.

(210) Keller, O. Electromagnetic Surface Waves on a Cooper-Paired Superconductor. *J. Opt. Soc. Am. B* **1990**.

(211) Glumova, M. V.; Lozovski, V. Z.; Reznik, D. V. Surface Waves on a Superconductor: Beyond the Weak-Coupling Approximation. *J. Phys. Condens. Matter* **2002**.

(212) Tsiatmas, A.; Buckingham, A. R.; Fedotov, V. A.; Wang, S.; Chen, Y.; De Groot, P. A. J.; Zheludev, N. I. Superconducting Plasmonics and Extraordinary Transmission. *Appl. Phys. Lett.* **2010**.

(213) Tsiatmas, A.; Fedotov, V. A.; García De Abajo, F. J.; Zheludev, N. I. Low-Loss Terahertz Superconducting Plasmonics. *New J. Phys.* **2012**.

(214) Chen, X.; Hu, D.; Mescall, R.; You, G.; Basov, D. N.; Dai, Q.; Liu, M. Modern Scattering-Type Scanning Near-Field Optical Microscopy for Advanced Material Research. *Advanced Materials*. 2019.

(215) Pimenov, A.; Engelbrecht, S.; Shuvaev, A. M.; Jin, B. B.; Wu, P. H.; Xu, B.; Cao, L. X.; Schachinger, E. Terahertz Conductivity in FeSe0.5Te0.5 Superconducting Films. *New J. Phys.* **2013**.

(216) Brorson, S. D.; Buhleier, R.; Trofimov, I. E.; White, J. O.; Ludwig, C.; Balakirev, F. F.; Habermeier, H.-U.; Kuhl, J. Electrodynamics of High-Temperature Superconductors Investigated with Coherent Terahertz Pulse Spectroscopy. *J. Opt. Soc. Am. B* **1996**.

(217) Ryusuke, M.; Naoto, T.; Hiroyuki, F.; Arata, S.; Kazumasa, M.; Yoshinori, U.; Hirotaka, T.; Zhen, W.; Hideo, A.; Ryo, S. Light-Induced Collective Pseudospin Precession Resonating with Higgs Mode in a Superconductor. *Science (80-. ).* **2014**.

(218) Wunsch, B.; Stauber, T.; Sols, F.; Guinea, F. Dynamical Polarization of Graphene at Finite Doping. *New J. Phys.* **2006**, *8* (12), 318–318.

(219) Hwang, E. H.; Das Sarma, S. Dielectric Function, Screening, and Plasmons in Two-Dimensional Graphene. *Phys. Rev. B* **2007**, *75* (20), 205418.

(220) Katsumi, K.; Tsuji, N.; Hamada, Y. I.; Matsunaga, R.; Schneeloch, J.; Zhong, R. D.; Gu, G. D.; Aoki, H.; Gallais, Y.; Shimano, R. Higgs Mode in the $d$-Wave Superconductor $Bi_2Sr_2$. *Phys. Rev. Lett.* **2018**, *120*

(11), 117001.
(221) Buzzi, M.; Jotzu, G.; Cavalleri, A.; Cirac, J. I.; Demler, E. A.; Halperin, B. I.; Lukin, M. D.; Shi, T.; Wang, Y.; Podolsky, D. Higgs-Mediated Optical Amplification in a Non-Equilibrium Superconductor. **2019**.
(222) Mounet, N.; Gibertini, M.; Schwaller, P.; Campi, D.; Merkys, A.; Marrazzo, A.; Sohier, T.; Castelli, I. E.; Cepellotti, A.; Pizzi, G.; et al. Two-Dimensional Materials from High-Throughput Computational Exfoliation of Experimentally Known Compounds. *Nat. Nanotechnol.* **2018**.
(223) Menichetti, G.; Calandra, M.; Polini, M. Electronic Structure and Magnetic Properties of Few-Layer Cr2Ge2Te6: The Key Role of Nonlocal Electron-Electron Interaction Effects. *2D Mater.* **2019**.
(224) Costa, A. T.; Santos, D. L. R.; Peres, N. M. R.; Fernández-Rossier, J. Topological Magnons in CrI$_3$ Monolayers: An Itinerant Fermion Description. **2020**.
(225) Costa, M. J. T.; Fernández-Rossier, J.; Peres, N. M. R.; Costa, A. T. Non-Reciprocal Magnons in a Two Dimensional Crystal with off-Plane Magnetization. **2020**.
(226) Xiao, D.; Chang, M. C.; Niu, Q. Berry Phase Effects on Electronic Properties. *Rev. Mod. Phys.* **2010**.
(227) Nagaosa, N.; Sinova, J.; Onoda, S.; MacDonald, A. H.; Ong, N. P. Anomalous Hall Effect. *Rev. Mod. Phys.* **2010**.
(228) Song, J. C. W.; Rudner, M. S. Chiral Plasmons without Magnetic Field. *Proc. Natl. Acad. Sci. U. S. A.* **2016**.
(229) Kumar, A.; Nemilentsau, A.; Fung, K. H.; Hanson, G.; Fang, N. X.; Low, T. Chiral Plasmon in Gapped Dirac Systems. *Phys. Rev. B* **2016**.
(230) Mahoney, A. C.; Colless, J. I.; Peeters, L.; Pauka, S. J.; Fox, E. J.; Kou, X.; Pan, L.; Wang, K. L.; Goldhaber-Gordon, D.; Reilly, D. J. Zero-Field Edge Plasmons in a Magnetic Topological Insulator. *Nat. Commun.* **2017**.
(231) Haldane, F. D. M. Berry Curvature on the Fermi Surface: Anomalous Hall Effect as a Topological Fermi-Liquid Property. *Phys. Rev. Lett.* **2004**.
(232) Chen, J. Y.; Son, D. T. Berry Fermi Liquid Theory. *Ann. Phys. (N. Y).* **2017**.
(233) Gao, F.; Gao, Z.; Shi, X.; Yang, Z.; Lin, X.; Xu, H.; Joannopoulos, J. D.; Soljačić, M.; Chen, H.; Lu, L.; et al. Probing Topological Protection Using a Designer Surface Plasmon Structure. *Nat. Commun.* **2016**, *7* (1), 11619.
(234) Khanikaev, A. B.; Shvets, G. Two-Dimensional Topological Photonics. *Nature Photonics*. 2017.
(235) Xu, S. Y.; Belopolski, I.; Alidoust, N.; Neupane, M.; Bian, G.; Zhang, C.; Sankar, R.; Chang, G.; Yuan, Z.; Lee, C. C.; et al. Discovery of a Weyl Fermion Semimetal and Topological Fermi Arcs. *Science (80-. ).* **2015**.
(236) Song, J. C. W.; Rudner, M. S. Fermi Arc Plasmons in Weyl Semimetals. *Phys. Rev. B* **2017**.
(237) Andolina, G. M.; Pellegrino, F. M. D.; Koppens, F. H. L.; Polini, M. Quantum Nonlocal Theory of Topological Fermi Arc Plasmons in Weyl Semimetals. *Phys. Rev. B* **2018**.
(238) Zyuzin, A. A.; Zyuzin, V. A. Chiral Electromagnetic Waves in Weyl Semimetals. *Phys. Rev. B - Condens. Matter Mater. Phys.* **2015**.
(239) Hofmann, J.; Das Sarma, S. Surface Plasmon Polaritons in Topological Weyl Semimetals. *Phys. Rev. B* **2016**.
(240) Hasdeo, E. H.; Song, J. C. W. Long-Lived Domain Wall Plasmons in Gapped Bilayer Graphene. *Nano Lett.* **2017**.
(241) Bliokh, K. Y.; Smirnova, D.; Nori, F. Quantum Spin Hall Effect of Light. *Science (80-. ).* **2015**.
(242) Jin, D.; Lu, L.; Wang, Z.; Fang, C.; Joannopoulos, J. D.; Soljačić, M.; Fu, L.; Fang, N. X. Topological Magnetoplasmon. *Nat. Commun.* **2016**.
(243) Jin, D.; Christensen, T.; Soljačić, M.; Fang, N. X.; Lu, L.; Zhang, X. Infrared Topological Plasmons in Graphene. *Phys. Rev. Lett.* **2017**.
(244) Shi, L. K.; Song, J. C. W. Plasmon Geometric Phase and Plasmon Hall Shift. *Phys. Rev. X* **2018**.
(245) Rudner, M. S.; Lindner, N. H. Band Structure Engineering and Non-Equilibrium Dynamics in Floquet Topological Insulators. *Nature Reviews Physics*. 2020.

(246) McIver, J. W.; Schulte, B.; Stein, F. U.; Matsuyama, T.; Jotzu, G.; Meier, G.; Cavalleri, A. Light-Induced Anomalous Hall Effect in Graphene. *Nature Physics*. 2020.
(247) Rudner, M. S.; Song, J. C. W. Self-Induced Berry Flux and Spontaneous Non-Equilibrium Magnetism. *Nat. Phys.* **2019**, *15* (10), 1017–1021.
(248) Singh, R.; Zheludev, N. Materials: Superconductor Photonics. *Nature Photonics*. 2014.
(249) Cao, Y.; Fatemi, V.; Fang, S.; Watanabe, K.; Taniguchi, T.; Kaxiras, E.; Jarillo-Herrero, P. Unconventional Superconductivity in Magic-Angle Graphene Superlattices. *Nature* **2018**.
(250) Lopes Dos Santos, J. M. B.; Peres, N. M. R.; Castro Neto, A. H. Graphene Bilayer with a Twist: Electronic Structure. *Phys. Rev. Lett.* **2007**.
(251) Bistritzer, R.; MacDonald, A. H. Moiré Bands in Twisted Double-Layer Graphene. *Proc. Natl. Acad. Sci. U. S. A.* **2011**.
(252) Lu, X.; Stepanov, P.; Yang, W.; Xie, M.; Aamir, M. A.; Das, I.; Urgell, C.; Watanabe, K.; Taniguchi, T.; Zhang, G.; et al. Superconductors, Orbital Magnets and Correlated States in Magic-Angle Bilayer Graphene. *Nature* **2019**.
(253) Stepanov, P.; Das, I.; Lu, X.; Fahimniya, A.; Watanabe, K.; Taniguchi, T.; Koppens, F. H. L.; Lischner, J.; Levitov, L.; Efetov, D. K. The Interplay of Insulating and Superconducting Orders in Magic-Angle Graphene Bilayers. **2019**.
(254) Rickhaus, P.; de Vries, F.; Zhu, J.; Portolés, E.; Zheng, G.; Masseroni, M.; Kurzmann, A.; Taniguchi, T.; Wantanabe, K.; MacDonald, A. H.; et al. Density-Wave States in Twisted Double-Bilayer Graphene. **2020**.
(255) Xie, M.; Macdonald, A. H. Nature of the Correlated Insulator States in Twisted Bilayer Graphene. *Phys. Rev. Lett.* **2020**.
(256) Stauber, T.; San-Jose, P.; Brey, L. Optical Conductivity, Drude Weight and Plasmons in Twisted Graphene Bilayers. *New J. Phys.* **2013**.
(257) Catarina, G.; Amorim, B.; Castro, E. V.; Castro, E. V.; Castro, E. V.; Lopes, J. M. V. P.; Lopes, J. M. V. P.; Peres, N. Twisted Bilayer Graphene: Low-Energy Physics, Electronic and Optical Properties. In *Handbook of Graphene*; 2019.
(258) Novelli, P.; Torre, I.; Koppens, F. H. L.; Taddei, F.; Polini, M. Optical and Plasmonic Properties of Twisted Bilayer Graphene: Impact of Interlayer Tunneling Asymmetry and Ground-State Charge Inhomogeneity. **2020**.
(259) Lewandowski, C.; Levitov, L. Intrinsically Undamped Plasmon Modes in Narrow Electron Bands. *Proc. Natl. Acad. Sci.* **2019**, *116* (42), 20869–20874.
(260) Calderón, M. J.; Bascones, E. Correlated States in Magic Angle Twisted Bilayer Graphene under the Optical Conductivity Scrutiny. **2019**.
(261) Sunku, S. S.; Ni, G. X.; Jiang, B. Y.; Yoo, H.; Sternbach, A.; McLeod, A. S.; Stauber, T.; Xiong, L.; Taniguchi, T.; Watanabe, K.; et al. Photonic Crystals for Nano-Light in Moiré Graphene Superlattices. *Science (80-. ).* **2018**.
(262) Hesp, N. C. H.; Torre, I.; Rodan-Legrain, D.; Novelli, P.; Cao, Y.; Carr, S.; Fang, S.; Stepanov, P.; Barcons-Ruiz, D.; Herzig-Sheinfux, H.; et al. Collective Excitations in Twisted Bilayer Graphene Close to the Magic Angle. **2019**.
(263) Akinwande, D.; Huyghebaert, C.; Wang, C.-H.; Serna, M. I.; Goossens, S.; Li, L.-J.; Wong, H.-S. P.; Koppens, F. H. L. Graphene and Two-Dimensional Materials for Silicon Technology. *Nature* **2019**, *573* (7775), 507–518.
(264) Kang, K.; Xie, S.; Huang, L.; Han, Y.; Huang, P. Y.; Mak, K. F.; Kim, C.-J.; Muller, D.; Park, J. High-Mobility Three-Atom-Thick Semiconducting Films with Wafer-Scale Homogeneity. *Nature* **2015**, *520* (7549), 656–660.
(265) Pezzini, S.; Mišeikis, V.; Piccinini, G.; Forti, S.; Pace, S.; Engelke, R.; Rossella, F.; Watanabe, K.; Taniguchi, T.; Kim, P.; et al. 30°-Twisted Bilayer Graphene Quasicrystals from Chemical Vapor Deposition. *Nano Lett.* **2020**, *20* (5), 3313–3319.